\documentclass[final,5p,times,twocolumn]{elsarticle} 
\usepackage{lineno,hyperref}
\modulolinenumbers[1]

\usepackage{gensymb}
\usepackage{graphicx}
\usepackage{subcaption}
\usepackage{amsmath}
\usepackage[capitalise]{cleveref}
\usepackage{upgreek}

\usepackage{pdfpages}
\usepackage{setspace}
\usepackage{comment}
\usepackage{placeins}
\usepackage{multirow}

\usepackage{siunitx}
\usepackage[version=4]{mhchem}

\usepackage{epstopdf}
\AppendGraphicsExtensions{.tif}

\usepackage{caption}
\begin{document}

\begin{frontmatter}

\title{Study of depth-dependent charge collection profiles in irradiated pad diodes}

\author[1]{M. Hajheidari \corref{cor1} }
\author[1]{M. Antonello}
\author[1]{E. Garutti}
\author[1]{R. Klanner}
\author[1]{J. Schwandt}
\author[1]{G. Steinbrück}

\address[1]{Institute for Experimental Physics, University of Hamburg, Luruper Chaussee 149, 22761 Hamburg, Germany}

\cortext[cor1]{Corresponding author: mohammadtaghi.hajheidari@desy.de}

\begin{abstract}
In this work, charge collection profiles of non-irradiated and irradiated \SI{150}{\um} thick $p$-type pad diodes were measured using a \SI{5.2}{GeV} electron beam traversing the diode parallel to the readout electrode. Four diodes were irradiated to \SI{1}{MeV} neutron equivalent fluences of 2, 4, 8, and \SI{12E15}{\per \square \cm} with \SI{23}{MeV} protons. The Charge Collection Efficiency profiles as a function of depth are extracted by unfolding the data. The results of the measurements are compared to the TCAD device simulation using three radiation damage models from literature which were tuned to different irradiation types and fluences. 
\end{abstract}


\end{frontmatter}
\section{Introduction}

\par For the upgrade of the High-Luminosity Large Hadron Collider (HL-LHC), the tracking detector in the CMS and ATLAS experiments will be exposed to the \SI{1}{MeV} neutron equivalent fluences up to \SI{2E16}{\per\square\cm} \cite{gianotti2005physics}. This level of fluences causes radiation damage in the silicon pixel sensors. The effects of the bulk damage are a change in the doping concentration, an increase in the depletion voltage, an increase in the leakage current, and an increase in the trapping rate of charge carriers \cite{8331152}. As a result, the Charge Collection Efficiency (CCE) is reduced which influences both the detection efficiency and the position reconstruction for segmented sensors. For understanding the performance of the pixel sensors as a function of the irradiation fluence, the CCE as a function of the position at which the charge carriers inside the sensor are generated needs to be known. 

\par The method for measuring the charge collection of a pad diode as a function of depth using an pion beam was introduced in \cite{gorivsek2016edge}. Later, in~\cite{HAJHEIDARI2022166177}, the first results of this measurements with electron beam were shown. As explained in \cite{HAJHEIDARI2022166177}, this method has several advantages over the conventional edge-TCT method with infrared light. Small angular spread of the beam inside the sensor (for a \SI{5.2}{GeV} electron beam, the estimated angular deflection in \SI{5}{mm} of silicon is \SI{0.53}{mrad}) which makes it possible to use the method for pad diodes. The collected charge can be normalised to absolute values as the average $dE/dx$ is known. The energy loss of the electron beam does not change after irradiation which is not the case for infrared light \cite{scharf2020influence}. 
\par The measured charge profile is the convolution of:
\begin{itemize}
    \item the CCE profile,
    \item the smearing caused by the limited spatial resolution of the measurements, 
    \item the energy deposition profile of the electron beam as a function of depth in the diode.
\end{itemize}

\par To unfold the measured data and obtain the CCE profile, the effects of the limited spatial resolution of the beam and the energy deposition profile should be taken into account.  

In this work, the results of edge-on measurements with an electron beam for two non-irradiated and four irradiated pad diodes are shown. The measurement results are then compared with damage models from literature: Hamburg Penta Trap Model (HPTM) \cite{8824412} and two models from the University of Perugia~\cite{morozzi2019tcad,Croci_2022}. Finally, the CCE corrected for the finite beam resolution and the energy deposition profile are presented. 

\section{Measurement Setup}
\label{section::edgeon::setup}
\begin{figure}[htb]
    \centering
	\includegraphics[width=.5\textwidth]{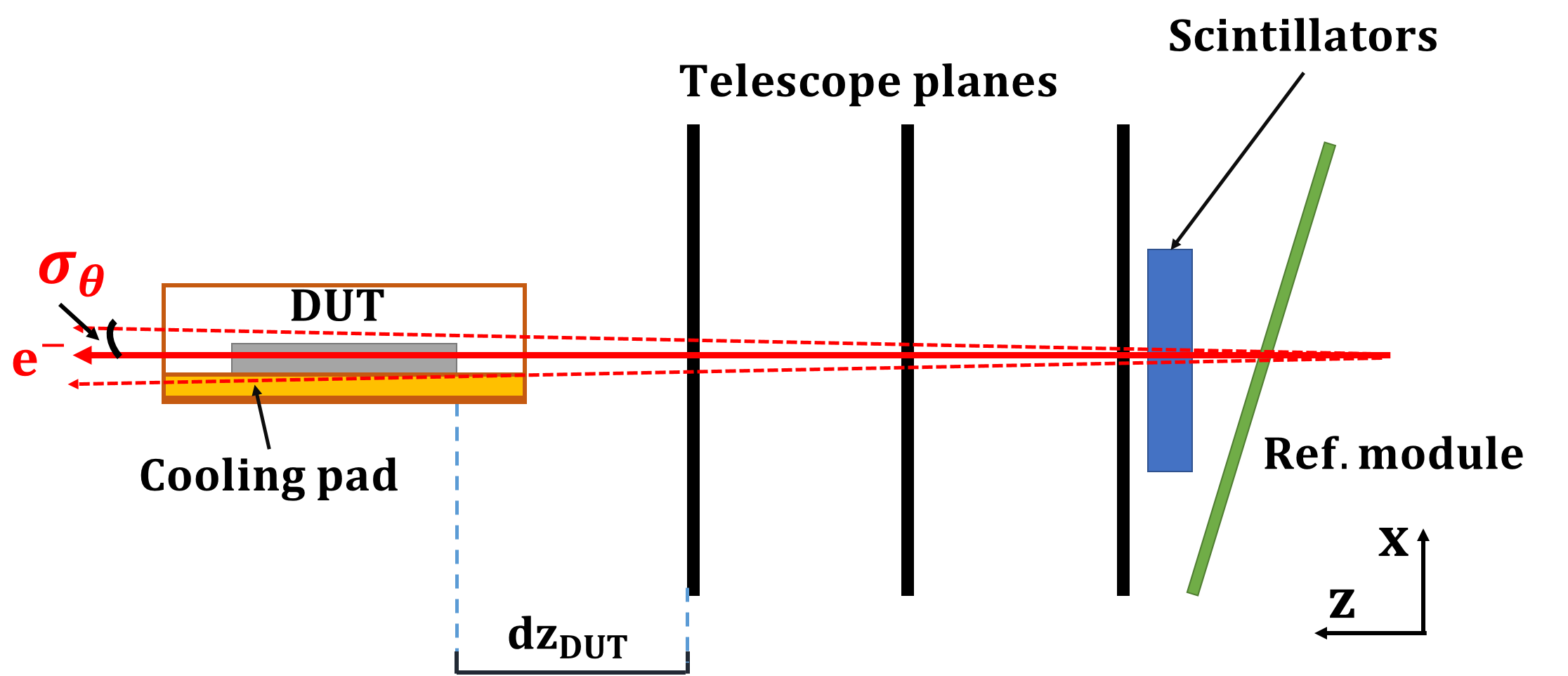} 
	\caption{Schematic of the measurement setup, taken from \cite{HAJHEIDARI2022166177}.}
     \label{setup}
\end{figure}

The measurements were carried out in the DESY II test-beam facility \cite{diener2019desy} with an electron beam with an energy of \SI{5.2}{GeV}. A schematic of the measurement setup is shown in \cref{setup}. 
\par For track reconstruction, three planes of the beam telescope called "DATURA" were used \cite{jansen2016performance}, as shown in \cref{setup}. Each plane of the telescope is equipped with a MIMOSA 26 sensor with a pixel size of \SI{18.4 x 18.4}{\um} and a thickness of \SI{54.5}{\um} \cite{jansen2016performance}. The readout of the telescope is binary with a threshold that corresponds to 6 times the pixel noise \cite{hu2010first}. The single hit resolution of each plane at vertical incidence is \SI{3.24}{\um}. 

\par The memory time of the telescope is relatively long (\SI{115}{\us}). To select the subset of the tracks which are in the readout cycle of the DUT, a timing reference module was used, as shown in \cref{setup}. This module is a CMS Phase-1 pixel module with a pixel size of \SI{100 x 150}{\um} and a thickness of \SI{285}{\um} \cite{Adam_2021}. The module has a readout rate of \SI{40}{\MHz} which corresponds to the readout cycle of \SI{25}{\ns}.

\par The trigger for the readout of the telescope, the time reference module, and the DUT was provided by the coincidence of the signals of two scintillators which were mounted in front of the telescope planes. The triggering area using the two scintillators was \SI{1.0 x 8.0}{\mm}.

\par
\begin{figure}[htb]
        \centering	\includegraphics[width=0.5\textwidth]{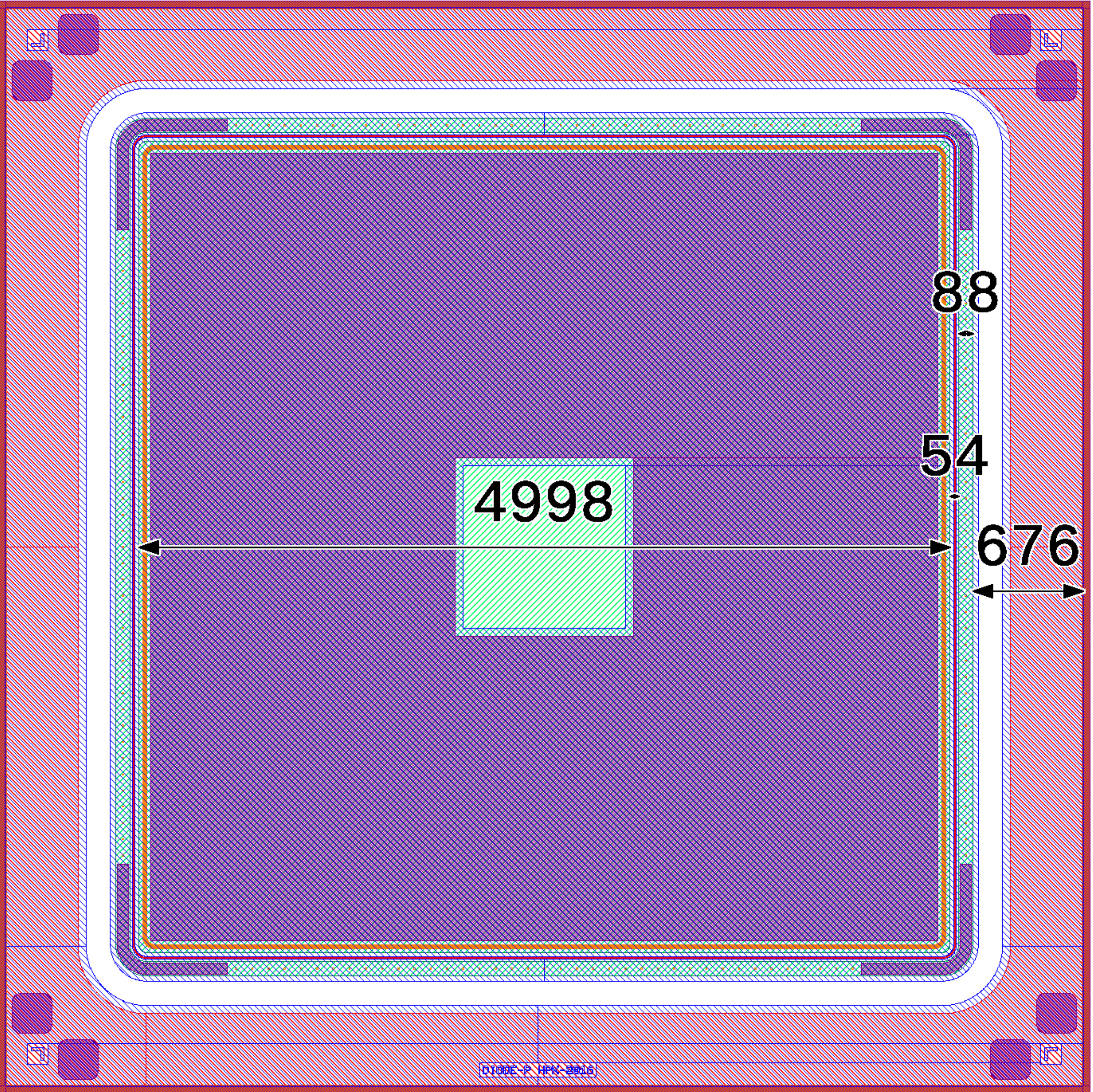} 
    \caption{Layout of the pad diode with a size of \SI{5 x 5}{\mm}. Dimensions are given in $\upmu \text{m}$.}
    \label{diode_top1}
\end{figure}

 The diodes investigated in this work are of $p$-type ($n^+pp^+$configuration) with a nominal thickness of \SI{150}{\um} produced by  Hamamatsu Photonics K.K (HPK). These diodes were produced as a part of the R\&D program for the Phase-2 upgrade of the inner tracker of the CMS detector. The diodes have an area of \SI{5.0 x 5.0}{\mm} and \SI{2.5 x 2.5}{\mm} with p-stop to isolate the pad from the guard ring electrically. \cref{diode_top1} shows the top view of the pad diode with a size of \SI{5 x 5}{\mm}. Four diodes were irradiated with \SI{23}{MeV} protons. Two diodes with the size of \SI{5.0 x 5.0}{\mm} were irradiated to \SI{1}{MeV} neutron equivalent fluences $\Phi_{\text{eq}}$ of \SI{2E15}{\per\square\cm} and \SI{4E15}{\per\square\cm}, and two diodes with the size of \SI{2.5 x 2.5}{\mm} were irradiated to $\Phi_{\text{eq}}$ of \SI{8E15}{\per\square\cm} and \SI{12E15}{\per\square\cm} at Karlsruhe Zyklotron AG \cite{ZAGwebsite}. High uniformity is reached by scanning the beam over the diodes. For determination of $\Phi_{\text{eq}}$, a hardness factor of $\kappa = 2.2$ is used \cite{allport2019experimental}. The irradiated sensors were stored at\SI{-28}{\degreeCelsius} except for the time they were irradiated, transported and handled for the measurements. From the capacitance-voltage measurement, the depletion voltage of the non-irradiated diode was determined to be around \SI{75}{V} and the doping density of the bulk region is about \SI{4.5E12}{cm^{-3}}. The guard ring of the diodes was floating during the measurement.

\par A Rohde \& Schwarz oscilloscope with an analog bandwidth of $4 \, \text{GHz}$ and a sampling rate of $20 \, \text{GS/s}$ recorded the transients \cite{RDosc}. A Femto HSA-X-40  amplifier with a bandwidth of $2.0 \, \text{GHz}$ and a nominal gain of $100$ was used to amplify the transients \cite{Femto}. In order to avoid the beam particle interacting in the cooling pad influencing the measurements, diodes and the cooling pad are separated by \SI{0.5}{mm} spacers.

To calculate the collected charge, $Q_{\text{meas}}$, the measured transients are integrated within a time window (gate). The charge is calculated as \cite{doi:10.1063/1.1710367}: 

\begin{equation}
{Q_{\text{meas}}}=\int_{{t}_0}^{{t}_1} \frac{{U(t)}}{G \cdot {R}_{{L}}} \ \text{d}t
\label{E1}
\end{equation}
\par $U(t)$ is the voltage transient after the baseline correction. $R_L$ is the input impedance of the oscilloscope (\SI{50}{\ohm}), $G$ is the nominal gain of the amplifier (100), and $t_0$ and $t_1$ are the start and end points of the integration gate. For this study, a gate width of \SI{30}{ns} was chosen and the time difference between the start of the gate, $t_0$ and the start of the pulse is \SI{10}{ns}. 

\par The following symbols are used throughout this paper: 
\begin{itemize}
    \item[] $d$: the thickness of the diode, 
    \item[] $x$: vertical distance from the center of the diode which extends from $-d/2$ ($n^+$ implant) to $+d/2$ ($p^+$ implant),
   \item[] $Q_{\text{meas}} (x)$: measured charge profile, 
    \item[] $CCE_e (x)$ and $CCE_h(x)$: Charge Collection Efficiency of electrons and holes as a function of $x$ using their charge collection lengths, i.e.\, $\lambda_{e,h}$ values, (\cref{eq::qeqh}),
    \item[] $CCE_{tot} (x)$: total Charge Collection Efficiency as a function of $x$ ($CCE_e (x)+CCE_h (x)$),
    \item[] $Q_{sim} (x)$: simulated charge profile using parameters from radiation damage models (\cref{eq::Qcal}),
    \item[] ${CCE}_{x_i}$: assumed Charge Collection Efficiency at $x_i$ used for unfolding,
    \item[] $CCE_{spl}(x)$: calculated Charge Collection Efficiency by spline interpolation between ${CCE}_{x_i}$ values,
    \item[] $Q_{sm}(x)$: calculated charge profile used for unfolding (\cref{eq::Qunf}).
\end{itemize}
\par In the analysis, the beam tracks are aligned with micrometre accuracy to the sensor coordinates by adding a "shift" to the beam-track position at the DUT in the $x$-direction. This shift is determined separately for the measurements of the different diodes. 
\par Before taking these data, the minimum tilt angle between the tracks and DUT was found using the online alignment procedure explained in \cite{HAJHEIDARI2022166177}. The analysis code for the beam track reconstruction is provided by Daniel Pitzl \cite{danielgithub}. In this analysis, the straight-track approximation is used.

\section{Results for the non-irradiated diodes}
\label{non-irrad}
\par
\begin{figure}[htb]
    \centering
	\includegraphics[width=.5\textwidth]{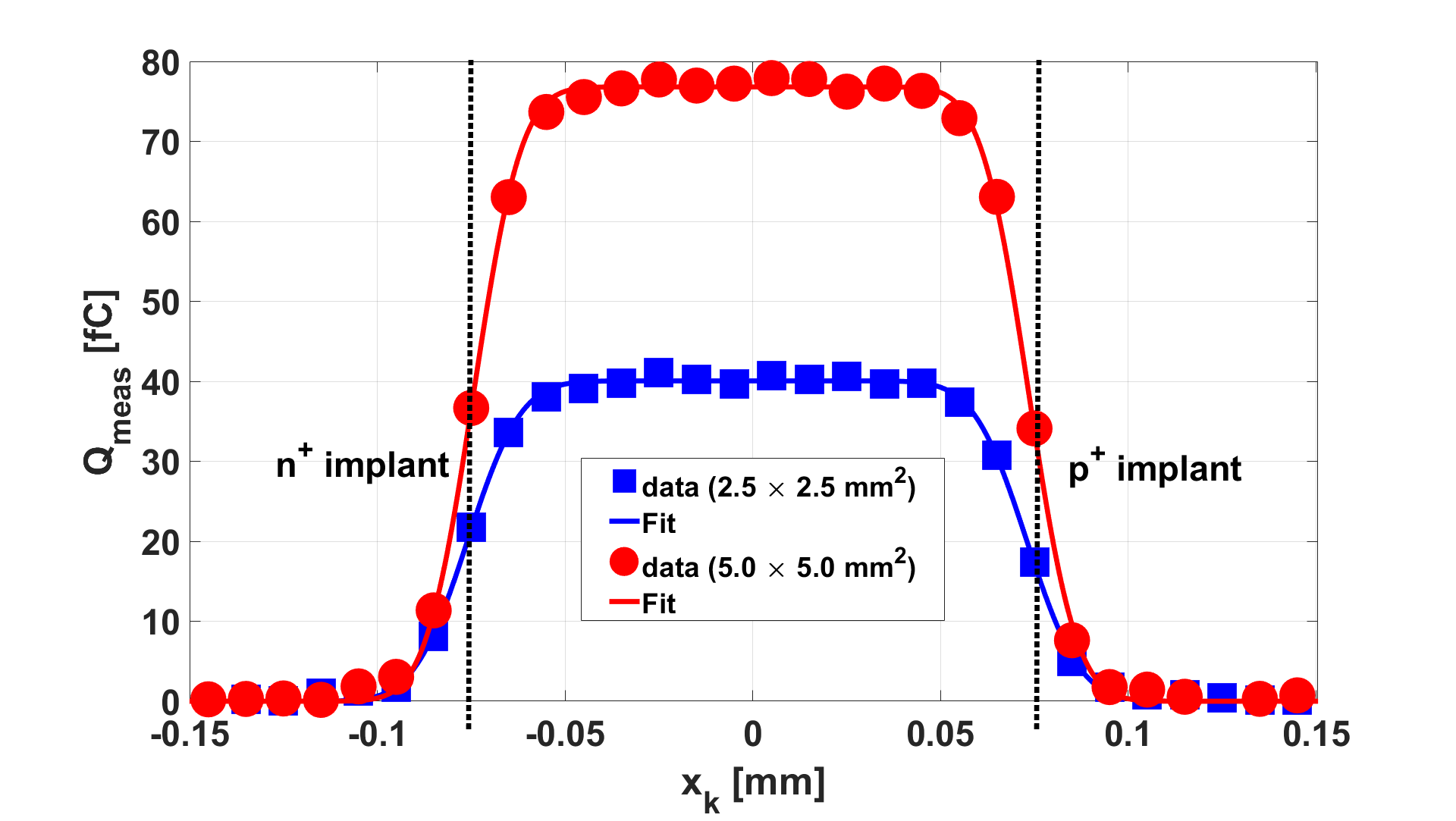} 
	\caption{Measured charge of the non-irradiated diodes as a function of $x_k$ at a bias voltage of \SI{120}{V}.}
     \label{qvsx_nonirrad}
\end{figure}

 The measured charge profiles of two non-irradiated diodes with sizes of \SI{2.5 x 2.5}{\per\square\cm} and \SI{5.0 x 5.0}{\per\square\cm} as a function of $x$ at a bias voltage of \SI{120}{V} are shown in \cref{qvsx_nonirrad}. The measurements were taken at room temperature. To exclude edge effects only the central region of the diode in the $y$-direction was selected. This region corresponds to $|y|<\SI{2.0}{\mm}$ and $|y|<\SI{1.0}{\mm}$ for diode sizes of \SI{5.0 x 5.0}{\square \cm} and \SI{2.5 x 2.5}{\square \cm}, respectively.  

\par As expected, for the non-irradiated diode, the profile is constant in the central region of the diode, i.e.\ $|x|<\SI{50}{\um}$ with an average value around \SI{77}{fC} and \SI{40}{fC} for the large and small diodes, respectively. At the edges, the collected charge is less due to two effects: 1. the smearing of the profiles because of the limited beam resolution, and 2. the loss at the edges due to energy leakage. To better understand the energy leakage, the mean energy deposition as a function of $x$ was simulated using the GEANT4 code \cite{AGOSTINELLI2003250}. 

\par
\begin{figure}[htb]

    \centering
	\includegraphics[width=0.5\textwidth]{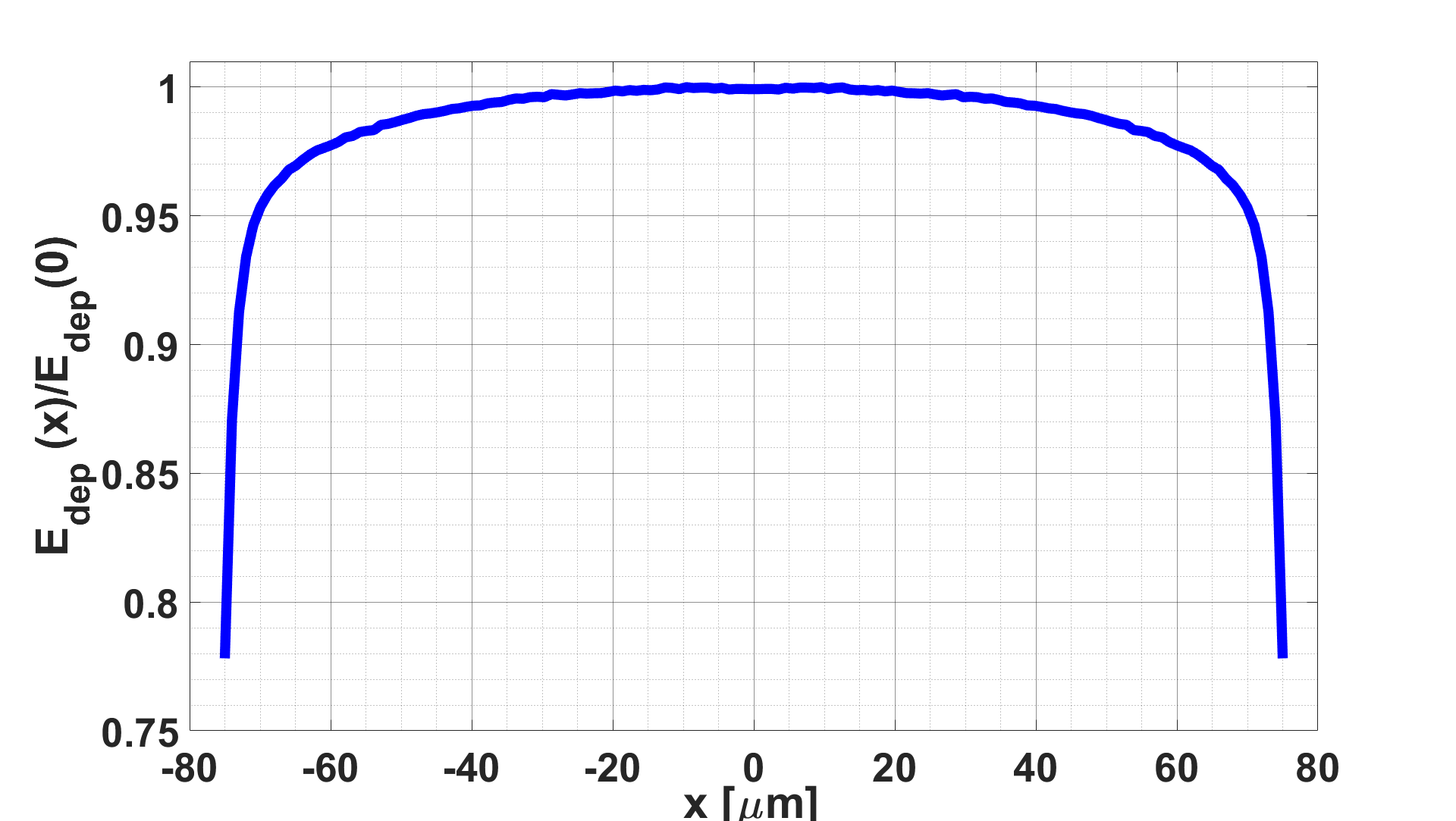} 
	\caption[Edge-on::Diode::Concept]{GEANT4 simulation of the relative energy deposition profile of \SI{5.0}{GeV} electron beam in \SI{5}{\mm} of silicon. The profile is normalised to its maximum value.}
	\label{geant4::edegeloss}
\end{figure}

 \cref{geant4::edegeloss} shows the result of the simulation of \SI{5.0}{GeV} electron tracks along \SI{5}{mm} of silicon. For this simulation, the \textit{PENELOPE} physics list was used \cite{sempau2003experimental}. The profile is normalised to its maximum value. One can see that the maximum energy deposition is at the centre ($x=\SI{0}{\um}$) and less energy is deposited near edges ($x=\pm \SI{75}{\um}$). The reason is that bremsstrahlung photons and pair-produced electrons and positrons have a higher chance of leaving the diode if the electron beam is close to the face of the diode.   

\par From the GEANT4 simulation, one can also estimate the absolute deposited energy and compare it with the data. The simulated deposited energy in center of the diode ($x=\SI{0}{\um}$) is \SI{1.767}{MeV}. By taking into account the elementary charge and ionisation energy in silicon (\SI{3.6}{eV}), the deposited charge is estimated as \SI{78.6}{fC} which is in agreement with the data at the\SI{2}{\%} level.   

\par To estimate the collected charge and the thickness of the non-irradiated diodes, the data is fitted with the following function:
\begin{equation}
\label{error_fit}
F(x) = \frac{A}{2}\cdot \left ( \text{erf} \Big(\frac{x-{\mu}_1}{\sqrt{2} \cdot \sigma}\Big)-\text{erf} \Big(\frac{x-{\mu}_2}{\sqrt{2}\cdot \sigma}\Big)\right)
\end{equation}

The free parameters of the fit are $A$, $\sigma$, $\mu_1$ and $\mu_2$. The parameter $A$ gives the scaling of the CCE profile and corresponds to the mean collected charge in the diode, $\mu_1$ and $\mu_2$ are the positions of the diode faces in test-beam coordinates, and $\sigma$ is the RMS width of the position resolution of the beam telescope. The results of the fit to the data are shown in \cref{qvsx_nonirrad} with solid lines. The fit is done in the range of \SIrange[]{-75}{+75}{\um}. The thickness of the diode is estimated as $\mu_2-\mu_1$. The results for the \SI{5 x 5}{\per\square\cm} and the \SI{2.5 x 2.5}{\per\square\cm} diodes are \SI{147.6 +- 0.4}{\um} and \SI{148.9 +- 0.6}{\um}. The RMS deviation between fit and data is \SI{0.74}{fC} and \SI{0.37}{fC} for large and small diodes, respectively. The spatial resolution of the beam telescope corresponds to $\sigma$ values which are \SI{10.6 \pm 0.2}{\um} and \SI{11.6 \pm 0.7}{\um}, for the large and the small diodes, respectively. 

\par
\begin{figure}[htb]
    \centering
	\includegraphics[width=.5\textwidth]{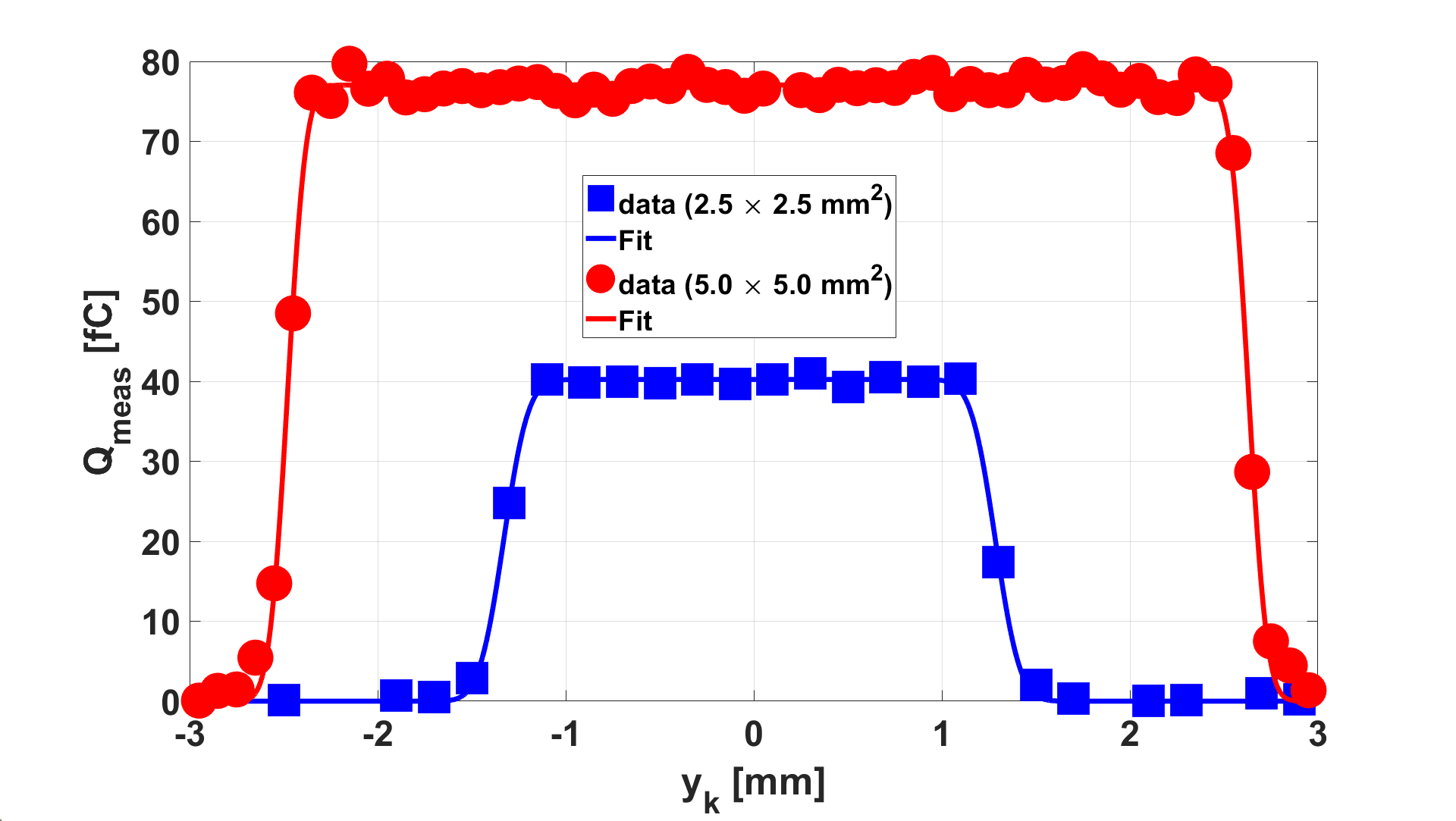} 
	\caption[DATURA]{Charge collection profiles of the non-irradiated diodes as a function of $y$. The profiles are taken for $|x|< \SI{50}{\um}$}. 
     \label{qvsy_nonirrad}
\end{figure}

 The ratio between the collected charge of the small to the large diode is:
\begin{align*} 
\frac{A_{small}}{A_{large}}=\frac{\SI{40.0+-0.2}{fC}}{\SI{76.8+-0.3}{fC}}=0.521 \pm 0.003
\end{align*}
Where $A_{small}$ and $A_{large}$ are the values obtained from the fit to the data for the \SI{2.5 x 2.5}{\mm} and the \SI{5.0 x 5.0}{\mm} diodes, respectively.
This ratio is higher than 0.50 which is the ratio of the widths of the implants. To check this number, one can estimate the width of the diodes from the charge profiles in the $y$-direction shown in \cref{qvsy_nonirrad}. Since the diodes are square-shaped in the $yz$-plane, the path length of the electron beam in the diode ($z$-direction) should be the same as the width of the diode ($y$-direction). Therefore, the ratio between the collected charges and the widths of the two diodes should be the same.
The profiles of \cref{qvsy_nonirrad} are fitted with the formula given in \cref{error_fit} and the widths of the diodes are determined as $\mu_2-\mu_1$. The ratio between the widths of two diodes ($W_{small}$ and $W_{large}$) is:
\begin{align*} 
\frac{W_{small}}{W_{large}}=\frac{\SI{2.64+-0.008}{mm}}{\SI{5.12+-0.004}{mm}}=0.515 \pm 0.001
\end{align*}

The results are similar to the ratio $A_{small}/A_{large}$. One sees that the estimated widths for both diodes are higher than the nominal values by $\approx \SI{130}{\um}$. This could be because the guard ring was floating during the measurements. Therefore, the charge is collected over a larger area than the pad area. 

\section{Results for the irradiated diodes}
\begin{figure}[htb]
 \begin{subfigure}[b]{1.0\linewidth}
    \centering	\includegraphics[width=1.0\textwidth]{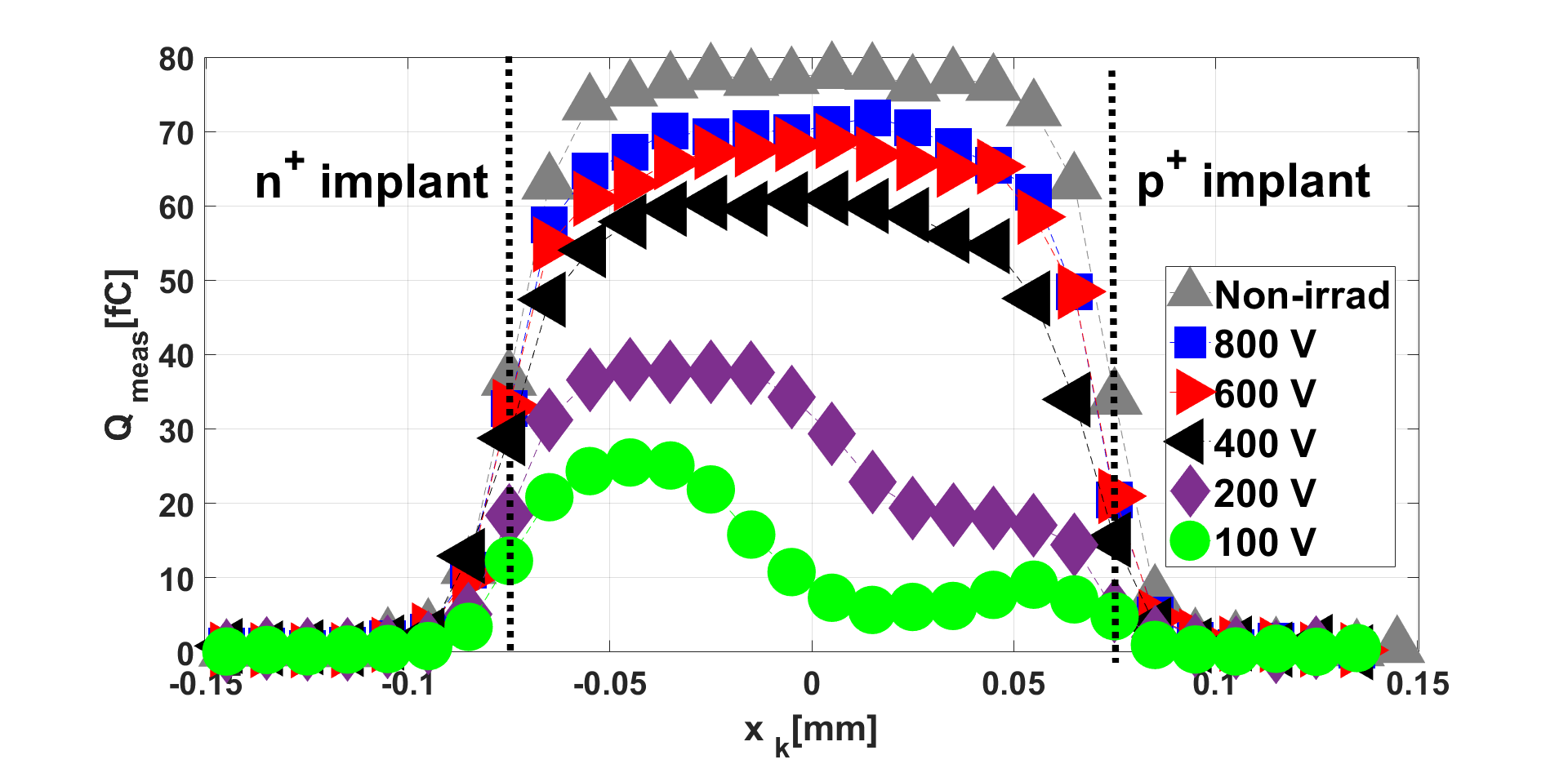} 
	\caption{$\Phi_{\text{eq}}=\SI{2E15}{\per\square\cm}$} 
     \label{2E15}
     \end{subfigure}
   \begin{subfigure}[b]{1\linewidth}
   \centering	\includegraphics[width=1\textwidth]{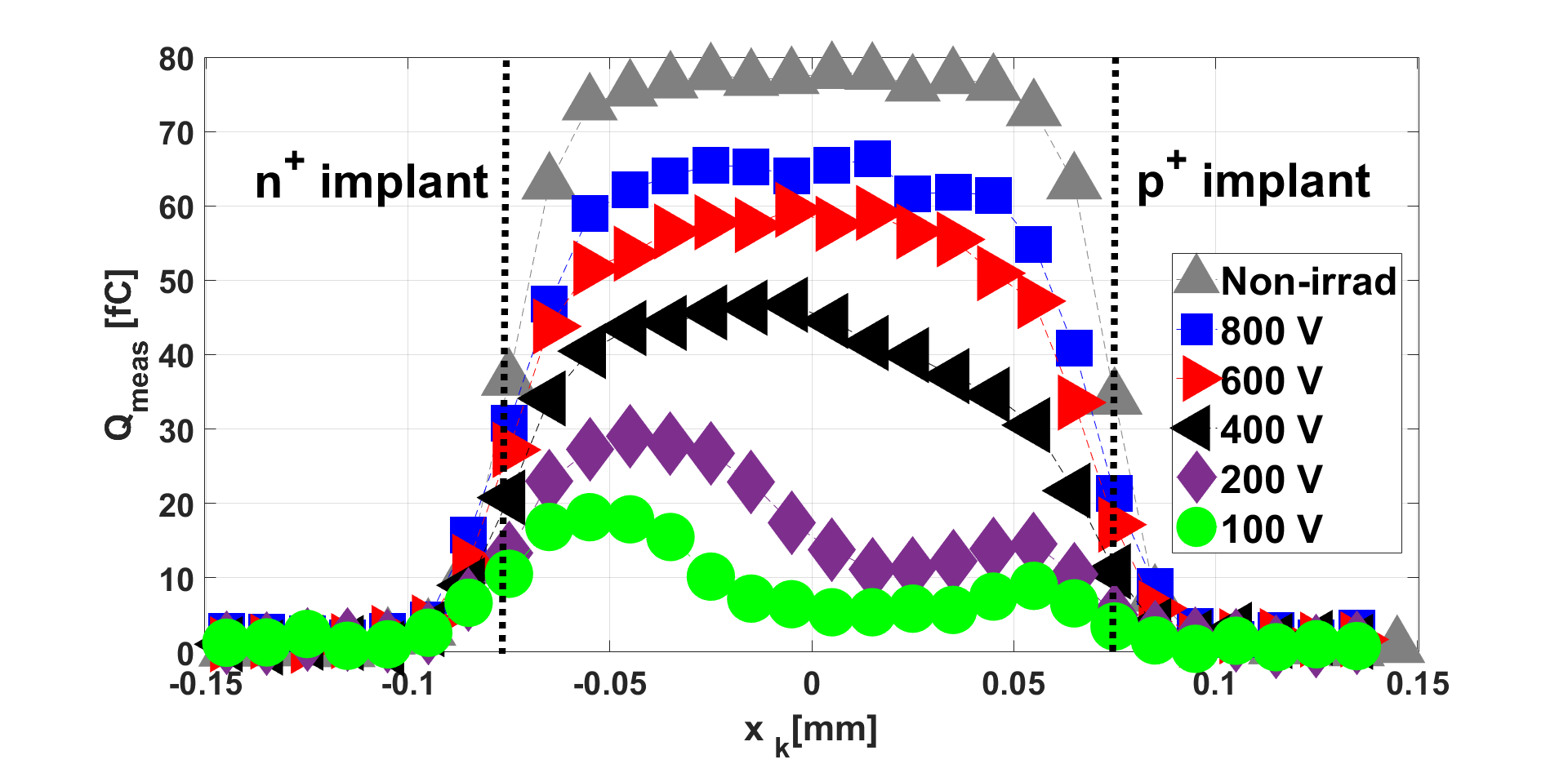} 
	\caption{$\Phi_{\text{eq}}=\SI{4E15}{\per\square\cm}$} 
	\label{4E15}
   \end{subfigure}
   \begin{subfigure}[b]{1\linewidth}
   \centering	\includegraphics[width=1\textwidth]{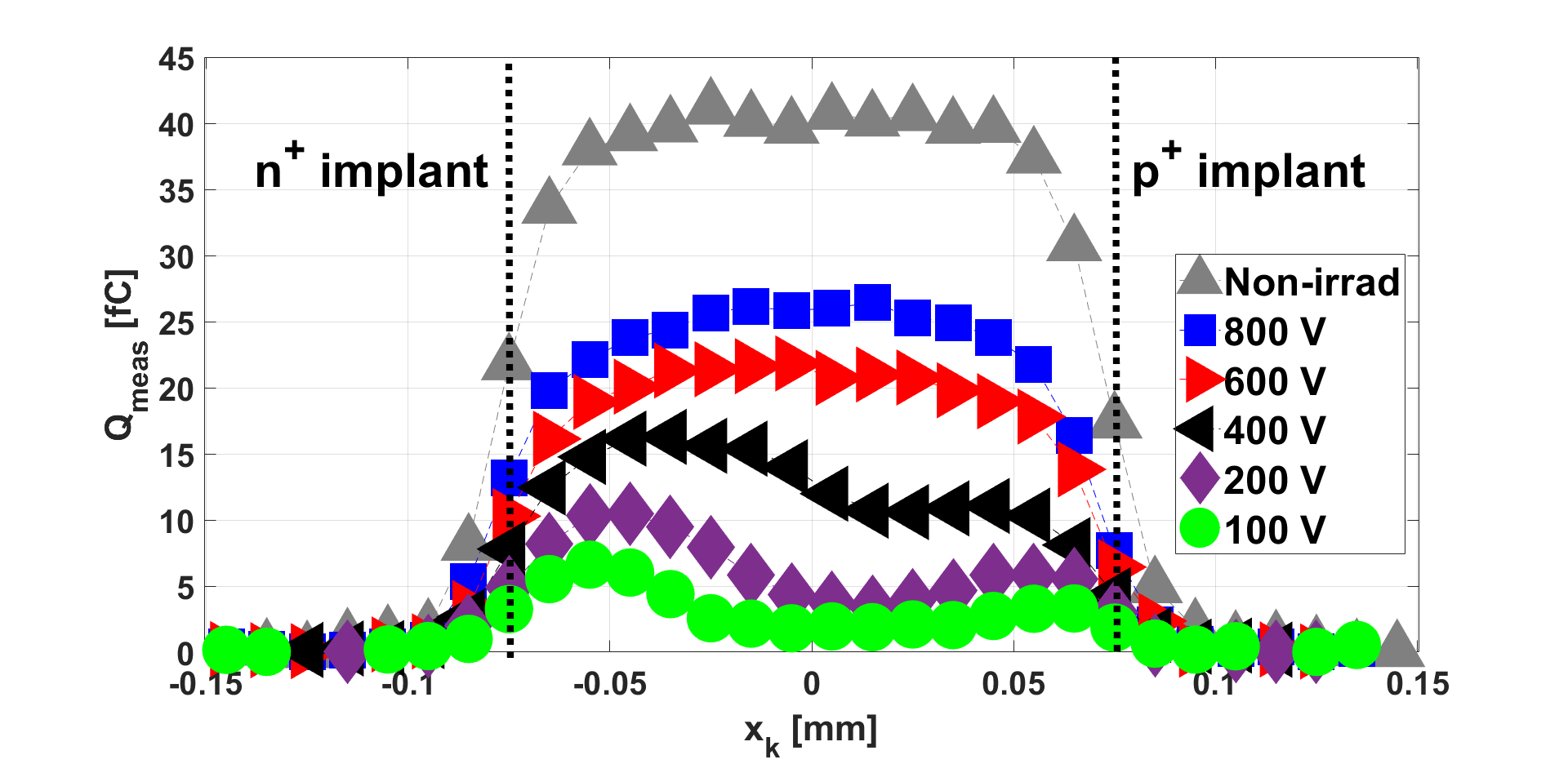} 
	\caption{$\Phi_{\text{eq}}=\SI{8E15}{\per\square\cm}$} 
	\label{8E15}
   \end{subfigure}
      \begin{subfigure}[b]{1\linewidth}
   \centering	\includegraphics[width=1\textwidth]{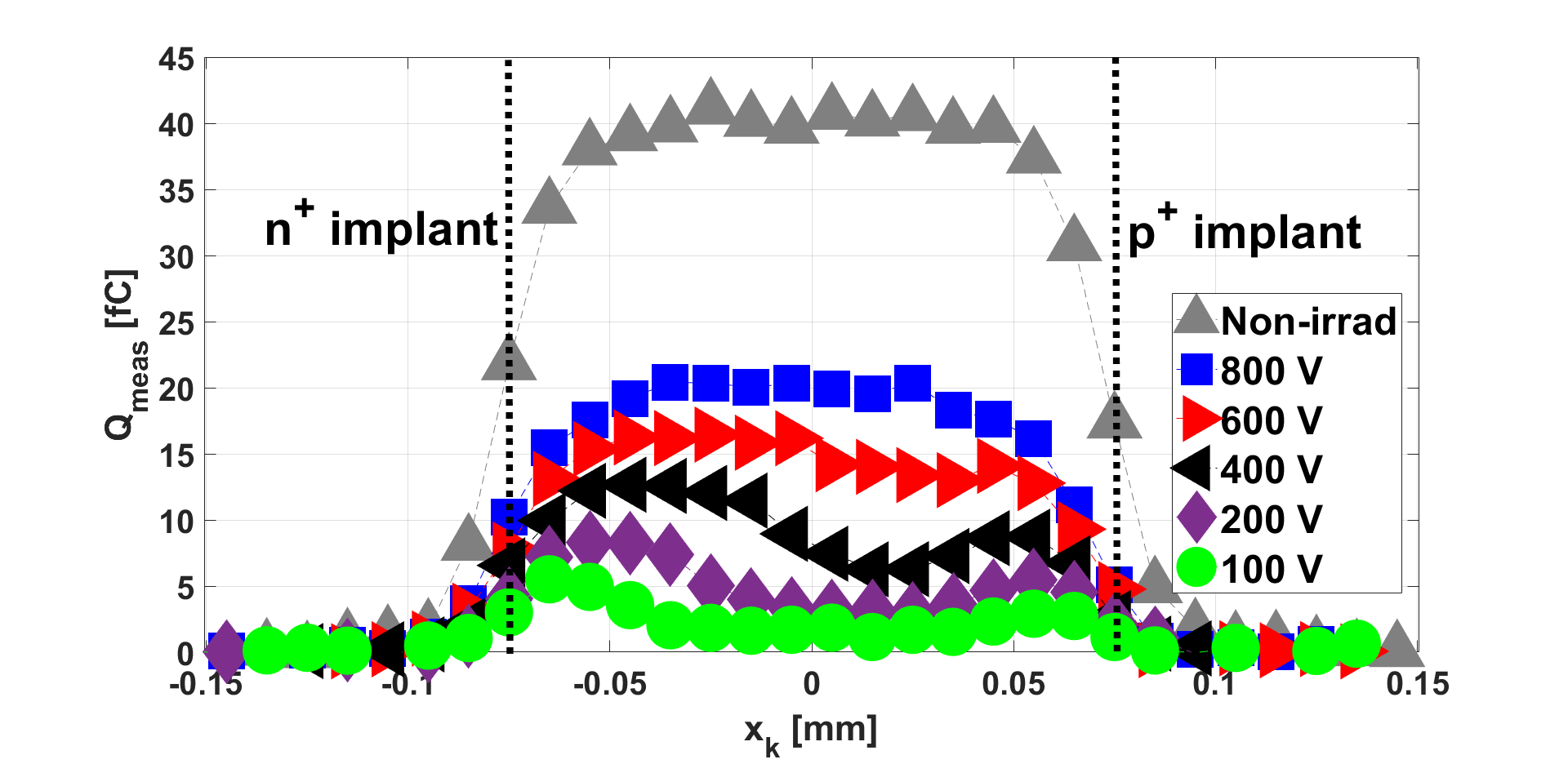} 
	\caption{$\Phi_{\text{eq}}=\SI{12E15}{\per\square\cm}$} 
	\label{12E15}
   \end{subfigure}
   \caption{Charge-collection profiles of the irradiated pad diodes at four irradiation fluences and different bias voltages. For each irradiated diode, in addition, the profile of the non-irradiated pad diode with the same size at a bias voltage of \SI{120}{V} is shown. The $n^+$ and $p^+$ implants are positioned at $x=-\SI{75}{um}$ and $x=\SI{+75}{\um}$, respectively (see \cref{qvsx_nonirrad})}.
   \label{qvsx_irr}
\end{figure}

The results of the measurements for the irradiated diodes at four irradiation fluences are shown in \cref{qvsx_irr}. For this data, the cold box was cooled down using a circulation chiller (operating at \SI{-20}{\degreeCelsius}) and two Peltier elements. The estimated value of the diode temperatures is \SI{-20 \pm 3}{\degreeCelsius}. For each diode, the measurements were taken at bias voltages between \SI{100}{V} and \SI{800}{V}. The in-situ alignment explained in \cite{HAJHEIDARI2022166177} was done at \SI{800}{V} for each diode. 
\par From these plots, the following observations are made:
\begin{itemize}
    \item the collected charge decreases at lower bias voltages and higher irradiation fluences,
    \item at high bias voltages, the charge profiles are uniform in the central region of the diode,
    \item at low bias voltages, the charge profiles are non-uniform and the region close to the $n^+$ implant has a higher charge collection than the region close to the $p^+$ implant with a minimum in the central region. 
\end{itemize}

\begin{figure}[htb]
 \begin{subfigure}[b]{1\linewidth}
    \centering	\includegraphics[width=1\textwidth]{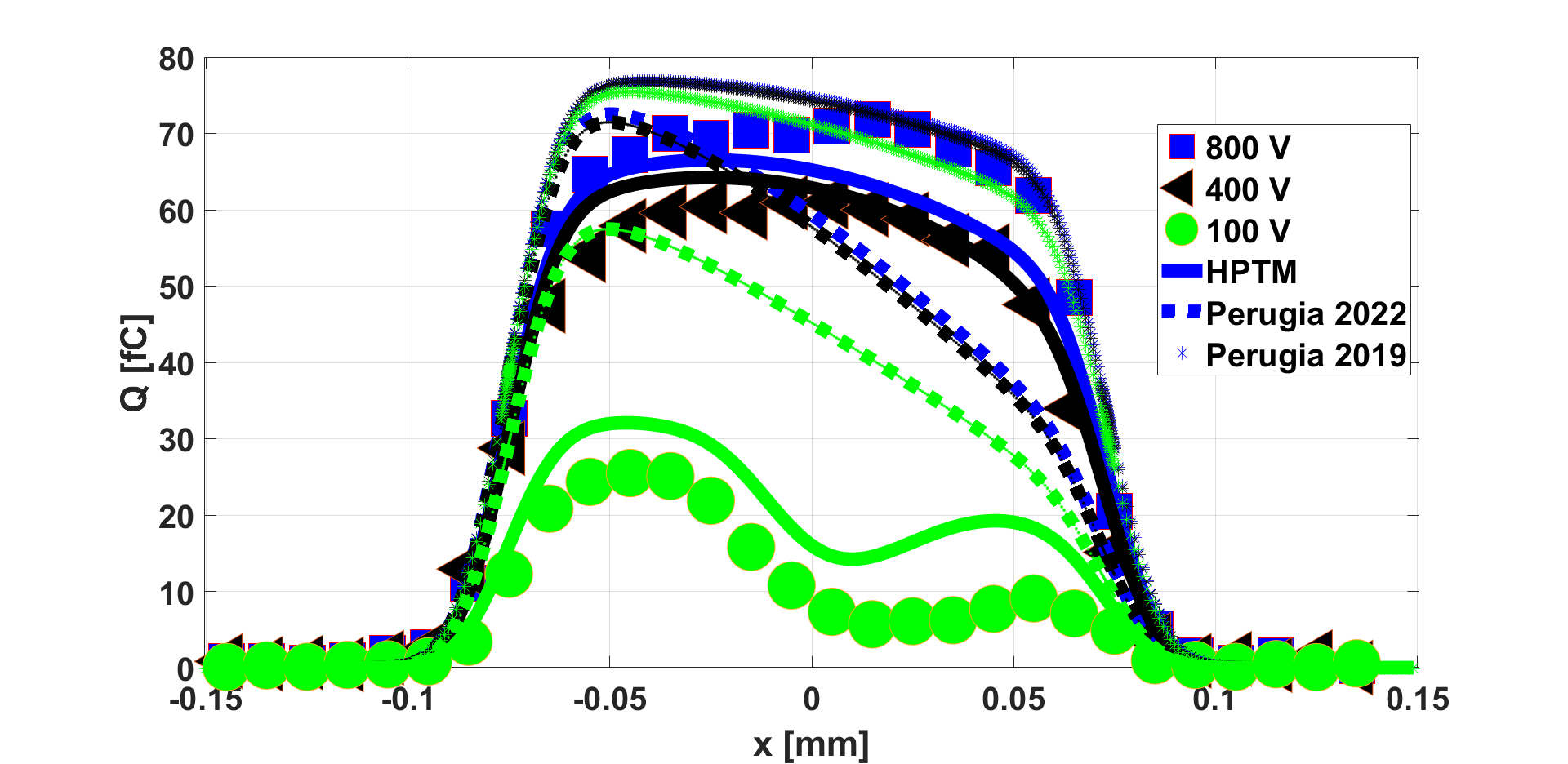} 
	\caption{$\Phi_{\text{eq}}=\SI{2E15}{\per\square\cm}$.} 
     \label{qvsx_2E15_simdata}
     \end{subfigure}
   \begin{subfigure}[b]{1\linewidth}
   \centering	\includegraphics[width=1\textwidth]{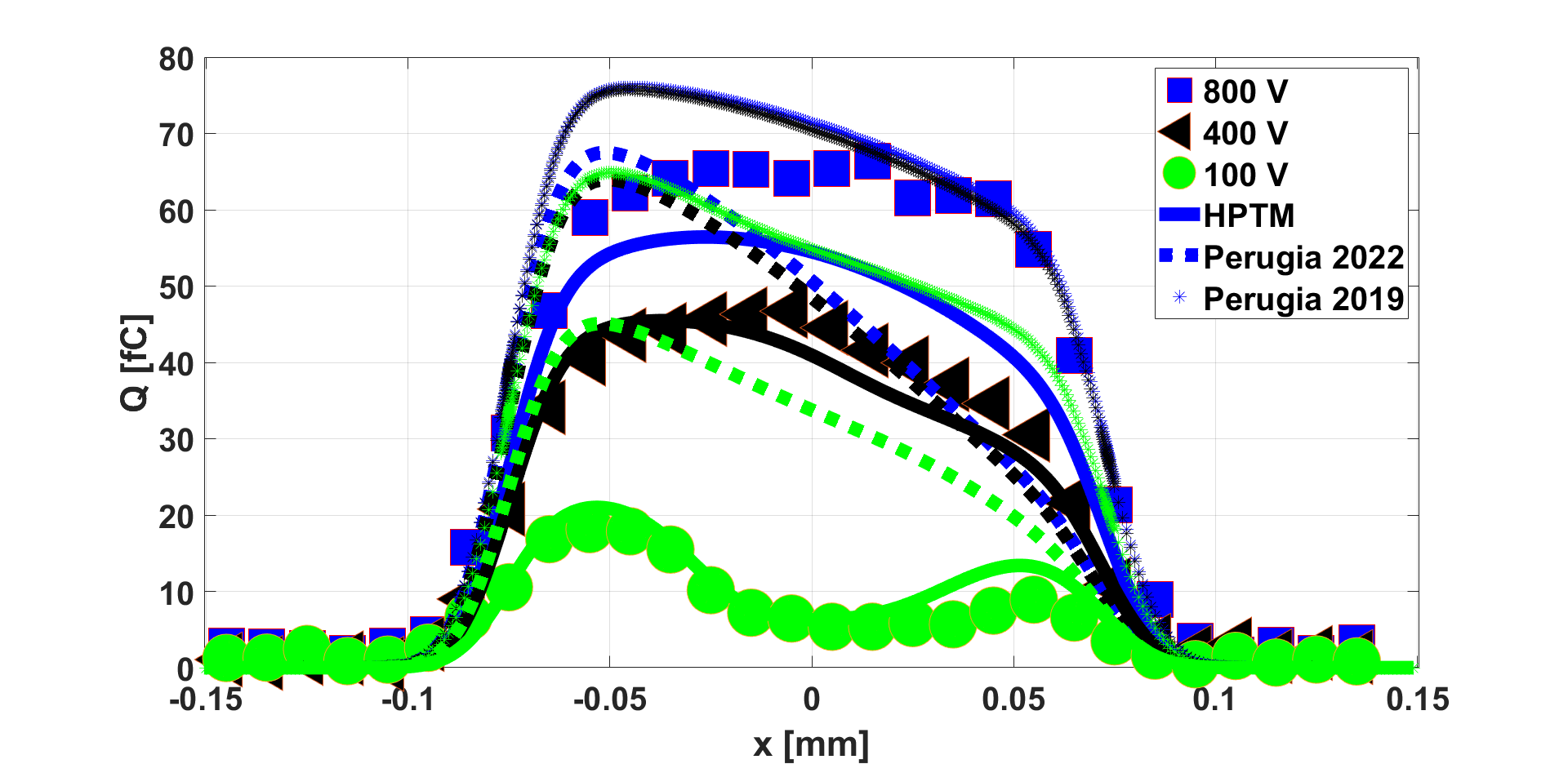} 
	\caption{$\Phi_{\text{eq}}=\SI{4E15}{\per\square\cm}$.} 
	\label{qvsx_4E15_simdata}
   \end{subfigure}
   \begin{subfigure}[b]{1\linewidth}
   \centering	\includegraphics[width=1\textwidth]{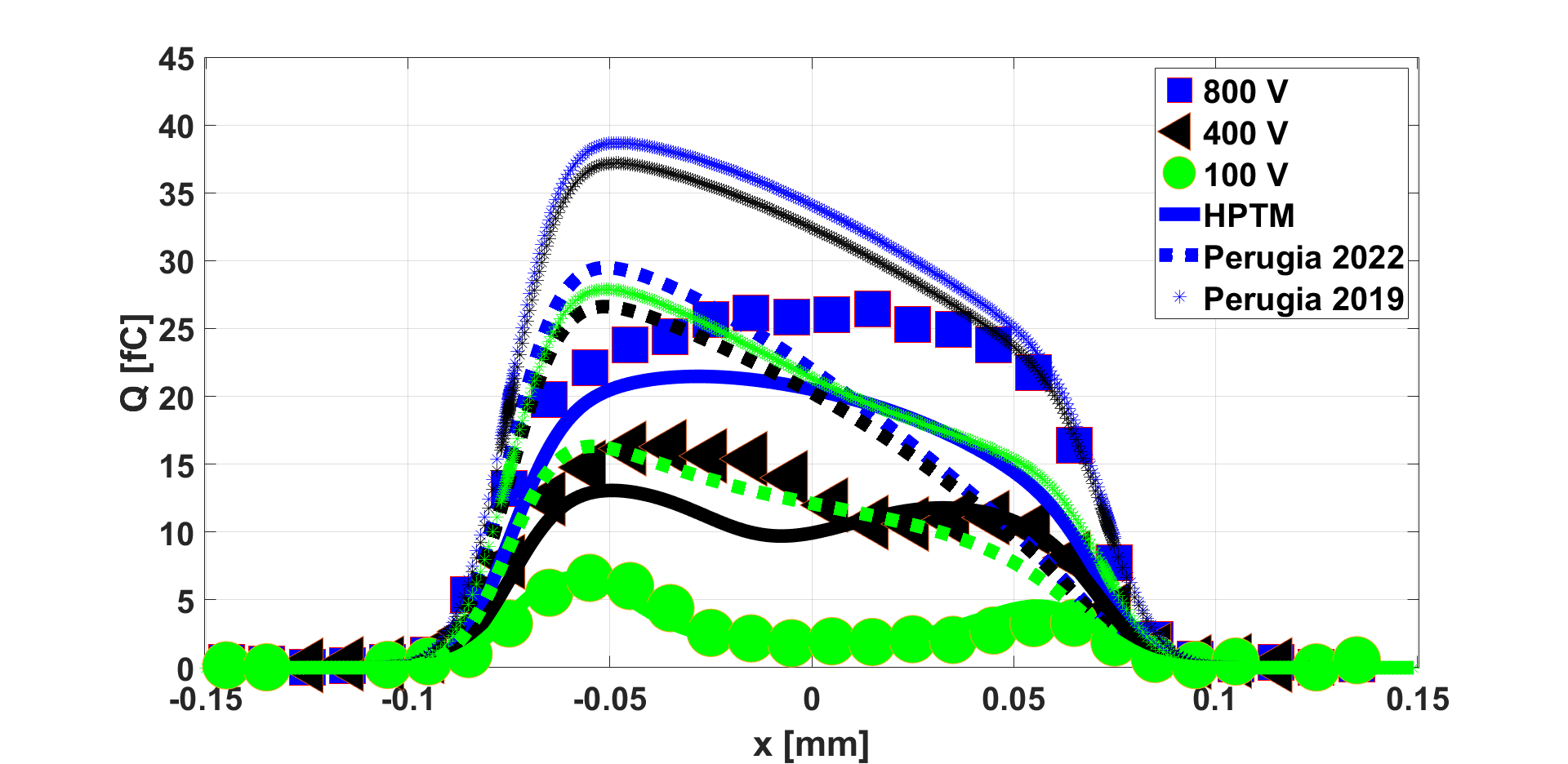} 
	\caption{$\Phi_{\text{eq}}=\SI{8E15}{\per\square\cm}$.} 
	\label{qvsx_8E15_simdata}
   \end{subfigure}
    \begin{subfigure}[b]{1\linewidth}
   \centering	\includegraphics[width=1\textwidth]{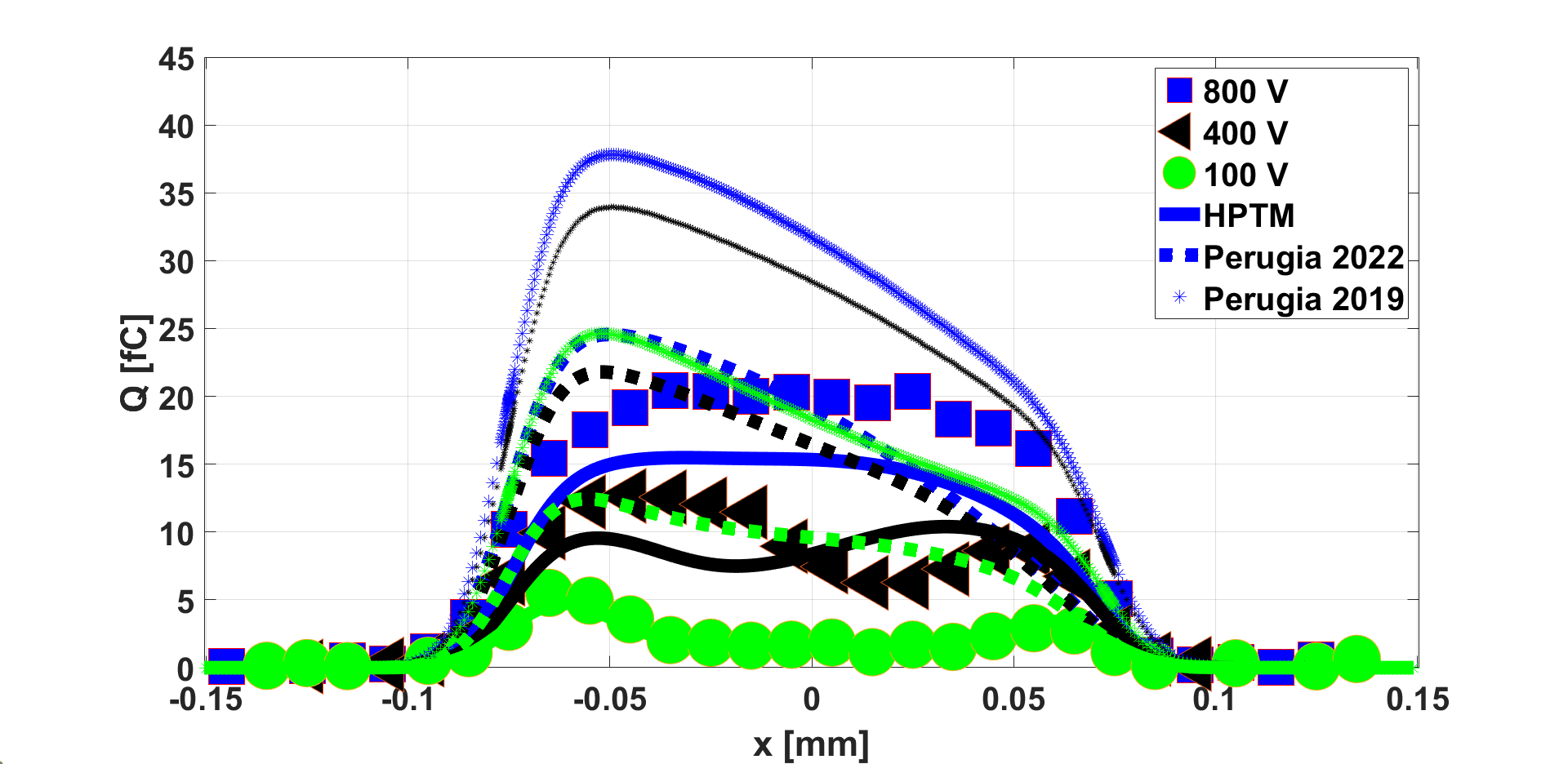} 
	\caption{$\Phi_{\text{eq}}=\SI{12E15}{\per\square\cm}$.} 
	\label{qvsx_12E15_simdata}
   \end{subfigure}
   
   \caption{Comparison between the charge-collection profiles and the results of the simulations at bias voltages of 800, 400 and \SI{100}{V} for each fluence. The data are shown with squares (\SI{800}{V}), triangles (\SI{400}{V}) and circles (\SI{100}{V}), and simulations are shown with solid (HPTM), dotted (Perugia 2022) and dashed (Perugia 2019) lines.}
   \label{ccevsx_data_tcad}
\end{figure}

\section{Comparison of the data with TCAD simulation}
\label{TCAD_HPTM}
\par In this section we show how the CCE profiles determined in this paper can be used to check simulations of radiation-damaged sensors. As examples, we compare the results to TCAD simulations using three published radiation damage models: the Hamburg Penta Trap Model (HPTM) and two Perugia models. However, the model parameters have been derived from measurements of sensors irradiated by different particles than the ones used in the present paper. The intention of this section is to show how the data can be used to test models for radiation-damaged silicon diodes.

\par The HPTM uses the optimiser of Synopsys TCAD \cite{TCAD}. It was introduced in \cite{8824412} to describe the experimental results on irradiated pad diodes. For that study, $p$-type diodes with \SI{200}{\um} thickness were irradiated at the CERN PS with \SI{24}{GeV/c} protons to various \SI{1}{MeV} neutron equivalent fluences in the range of \SIrange{0.3E15}{13E15}{\per \square \cm}. $I$-$V$, $C$-$V$, and $CCE$ measurements with Near Infra-Red (NIR) light of \SI{1065}{nm} were used for tuning the model. Moreover, the model does not take into account the reduction of the light transmission after irradiation \cite{scharf2020influence}. The HPTM assumes 5 defects in the silicon band-gap after irradiation. In \cite{8824412}, the type, energy levels, introduction rate, and the cross-section of each defect after optimisation are listed. 

\par The "New Perugia Model" was introduced in \cite{morozzi2019tcad}. The model assumes three defects (two acceptors and one donor) in the silicon band-gap after irradiation. For modelling the bulk damage, the model was compared with the results of charge collection for irradiated PiN diodes and $p$-type strip sensors. Charge collection of strip sensors irradiated with neutrons and \SI{26}{MeV} protons up to the equivalent fluence of \SI{2.2E16}{\per\square\cm} were measured with electrons from a \ce{^{90}_{}Sr} source at a bias voltage of \SI{900}{V} \cite{Affolder:2010zza}. PiN diodes irradiated with neutrons up to the equivalent fluence of \SI{1.0E16}{\per\square\cm} were measured with a NIR (\SI{1064}{nm}) pulsed laser and the results were normalised to the collected charge of the non-irradiated diode \cite{Ferrero2019}. A more recent version of the Perugia model was presented in \cite{Croci_2022}. The parameters of the model were optimised by comparing to the $I$-$V$ and $C$-$V$ of PiN and Low Gain Avalanche Diodes (LGADs). 

 \par From each model, using the optimised parameters, the position-dependent trapping times, $\tau_e (x)$ and $\tau_h (x)$, and drift velocities $v_e(x)$ and $v_h(x)$ are obtained. The Charge Collection Lengths ($\lambda_e$, $\lambda_h$ ) of electrons and holes are calculated as:
 \begin{equation}
\lambda_{e,h}(x)=v_{e,h}(x) \cdot \tau_{e,h}(x).
\label{eq::lambda}
\end{equation}
The $CCE$ of electrons and holes as a function of $x$ is obtained as:
 \begin{equation}
 \label{eq::qeqh}
 \begin{split}
    CCE_{e}(x) &=\int_{-d/2}^{x} {E_w(y)\cdot \text{exp}\left( \int_{x}^{y}{\frac{dy'}{\lambda_e(y')}} \right)} \ \text{d}y , \\
    CCE_{h}(x) &= \int_{x}^{+d/2} {E_w(y)\cdot \text{exp}\left( \int_{x}^{y}{-\frac{dy'}{\lambda_h(y')}} \right)} \ \text{d}y .
 \end{split}
\end{equation}
\par $E_w(y)$ is the weighting field which is $1/d$ for irradiated and non-irradiated pad diodes with a thickness of $d$ \cite{schwandt2019weighting}. The total CCE as a function of $x$ is the sum of the electron and hole contributions, i.e. $CCE_{tot}(x)=CCE_e(x)+CCE_h(x)$. To compare the simulation with measured charge profiles, $CCE_{tot}(x)$ has to be corrected for the energy leakage, the position resolution of the telescope and the normalisation as follows:
 \begin{equation}
Q_{sim}(x)= A \cdot \Big(CCE_{tot}(x) \cdot \frac{E_{dep}(x)}{E_{dep}(0)}\Big) * \text{Gauss} \, (x,0, \sigma) .
        \label{eq::Qcal}
\end{equation}
\par In this relation, $\frac{E_{dep} (x)}{E_{dep} (0)}$ is the profile shown in \cref{geant4::edegeloss} which takes into account the leakage effect; the convolution with $\text{Gauss} \, (x,0, \sigma)$ takes into account the limited spatial resolution of the beam telescope, and the multiplication with the parameter $A$ ,as determined in \cref{non-irrad}, scales the CCE profile. The convolution operator is shown with $*$. The assumed value for $\sigma$ is \SI{10}{\um} and the scaling factors, $A$, for the \SI{2.5 x 2.5}{\mm} and \SI{5.0 x 5.0}{\mm} diodes are \SI{40.0}{fC} and \SI{76.8}{fC}, respectively as determined in \cref{non-irrad}.

\par In \cref{ccevsx_data_tcad} a comparison between simulated and measured charge profiles is shown. The measured profile is compared to the simulated profiles from the HPTM and the two Perugia models (2019 and 2022). 

\par As already observed in \cite{8824412}, the simulated charge profiles at high bias voltages are lower than the measured ones. Furthermore, it should be noted that the HPTM was tuned to reproduce the results of the diodes irradiated with \SI{24}{GeV/c} protons, while the diodes in this work were irradiated with \SI{23}{MeV} protons. In \cite{Neubuser:2013yla}, it is shown that the CCE of irradiated diodes with \SI{23}{MeV} is higher than the diodes irradiated with \SI{24}{GeV/c} at a similar equivalent fluence. 

\par It is noted that none of the models can describe the observed position-, voltage- and fluence-dependence of the charge collection. This is not too surprising, as the models have been tuned using silicon sensors irradiated with different types of particles. However, the study demonstrates that the precise measurement of depth profiles provides stringent tests of models for the radiation damage of silicon detectors.

\section{Extracting CCE profiles from data}
\label{sec::ccevsx}
As discussed in \cref{non-irrad}, the measured charge profiles are affected by the limited beam position resolution and the energy-deposition profile. By unfolding the profiles, one can extract the $CCE$ profiles of pad diodes. 
\par The free parameters of the unfolding procedure are the $CCE_{xi}$ values at the 7 $x$ positions of $-65$, $-45$, $-25$, $0$, $25$, $45$, \SI{65}{\um}. ${CCE_{spl}}(x)$ is calculated by spline interpolation between ${CCE}_i$ values. For the interpolation, the "cubic spline interpolation" and the "not-a-knot" boundary condition were used. This condition means the third derivative of the interpolation function is continuous at the endpoints. For this calculation, MATLAB2019a was used (\cite{Matlab2019}). ${CCE}_{spl}(x)$ is multiplied with the energy deposition profile, $\frac{E_{dep}(x)}{E_{dep} (0)}$ shown in \cref{geant4::edegeloss}, convolved with a Gauss function with a standard deviation $\sigma$ and scaled with $A_{scale}$ calculated from the fit of the non-irradiated diode profiles with \cref{error_fit} (see \cref{eq::Qcal}). These steps can be written as:
\begin{equation}
Q_{sm}(x)= A \cdot \Big({CCE_{spl}}(x) \cdot \frac{E_{dep}(x)}{E_{dep} (0)}\Big) * \text{Gauss} \, (x,0, \sigma).
        \label{eq::Qunf}
\end{equation}
The values of ${CCE}_{i}$ are obtained by minimising the following function: 
\begin{equation}
\begin{split}
    D^2 & =  \sum_{k} {\Big({Q_{meas,k}-Q_{sm}(x_k)}\Big)^2}+\\ 
  & w_{pen}\sum_{i=2}^{6}{\left( \frac{0.5\cdot({CCE}_{i-1}+{CCE}_{i+1})-{CCE}_{i}}{\Delta x}\right)}^2 .  
\end{split}
   \end{equation}

\par In this function, $Q_{\text{meas},k}$ and $Q_{sm}(x_k)$ are the measured and calculated  charge at $x_k$, respectively.

\par The first term in the above function minimises the differences between the data and the model calculation for the $n_k$ measurements. The second term is the second derivative of the ${CCE}_{x_i}$ values. By minimising this term, the ${CCE}_{x_i}$ profile is smoothed. The penalty weight, $w_{pen}$, is adjusted for each profile separately depending on the shape of the profile. At high-bias voltages with more uniform charge collection profiles, $w_{pen}$ is chosen to be around 1. As the charge collection profiles become less uniform at lower bias voltages and higher fluences, lower value for $w_{pen}$ are chosen. 

\par In this procedure, the value of $\sigma$ is adjusted for each profile, separately. For each profile, $\sigma$ is changed manually in steps of \SI{0.5}{\um} in the range of \SIrange{9.0}{11.0}{\um} to find the minimum value for $D^2$. The same is done for the value of the shift parameter with steps of \SI{1}{\um}. The fit range is set at \SIrange{-75}{+75}{\um}. The assumed thickness of the diodes is \SI{150}{\um}. 

\par
\begin{figure}[htb]
 \begin{subfigure}[b]{.9\linewidth}
    \centering	\includegraphics[width=1\textwidth]{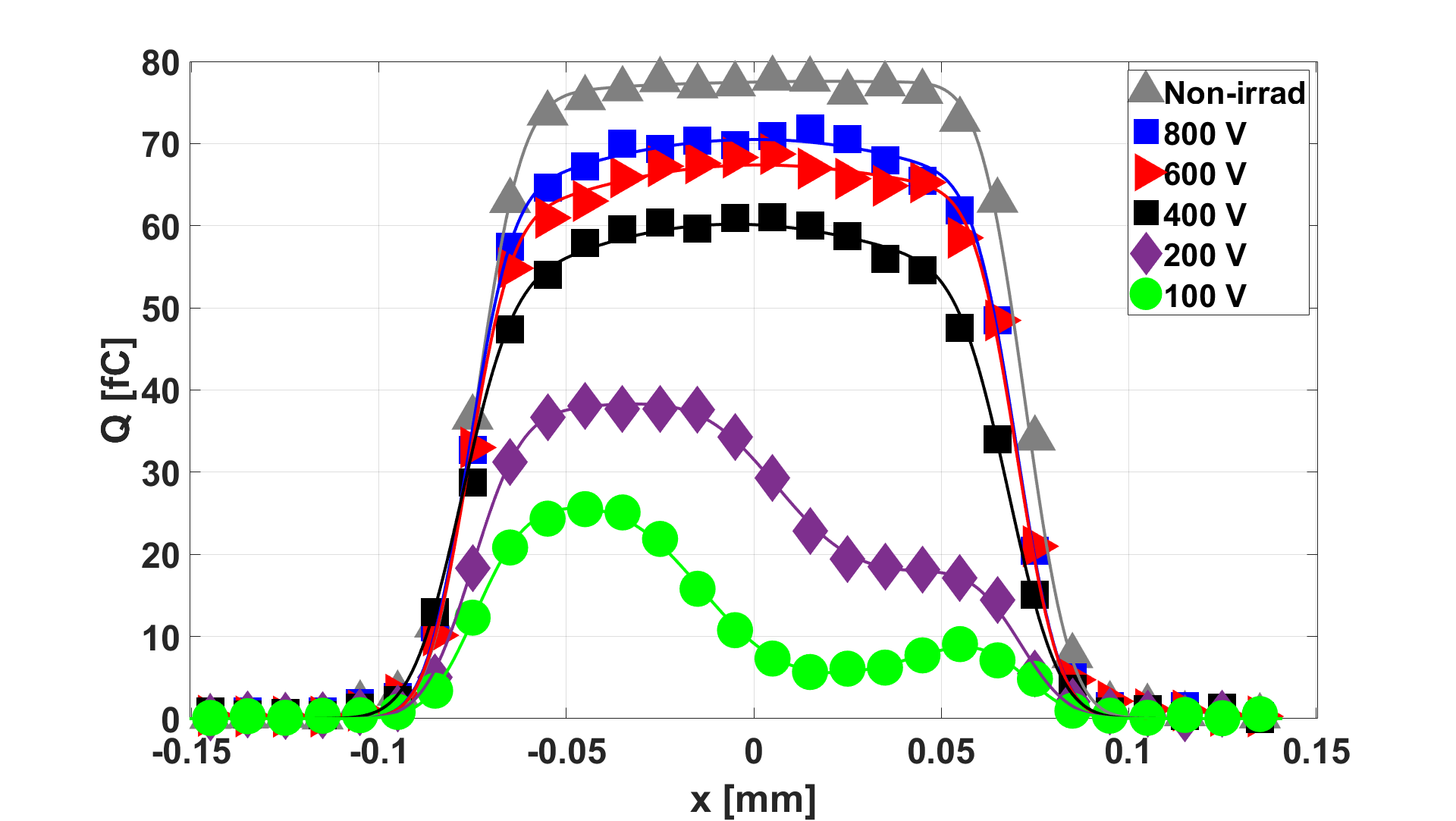} 
	\caption{$\Phi_{\text{eq}}=\SI{2E15}{\per\square\cm}$} 
     \label{2E15_data_fit}
     \end{subfigure}
   \begin{subfigure}[b]{.9\linewidth}
   \centering	\includegraphics[width=1\textwidth]{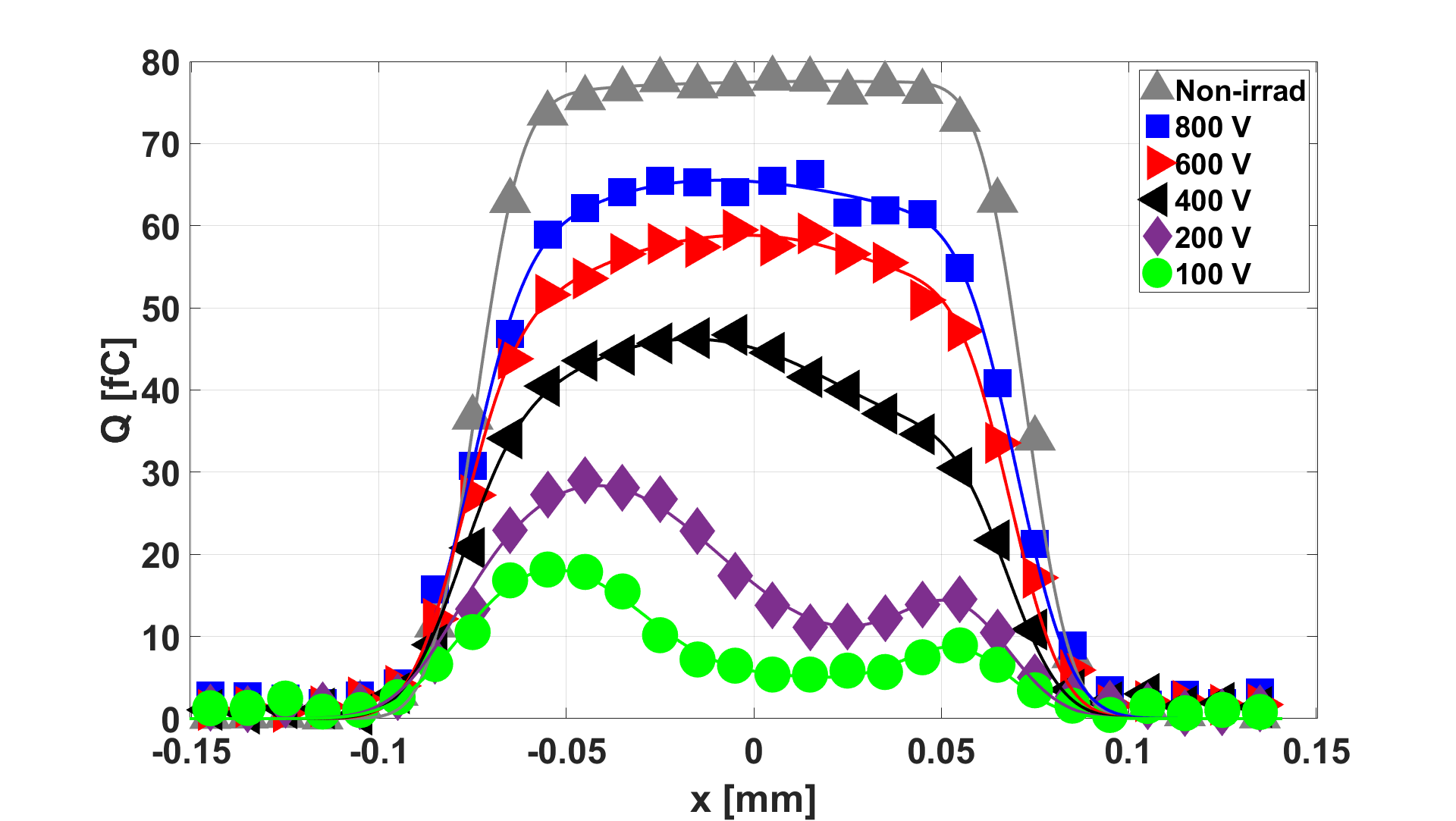} 
	\caption{$\Phi_{\text{eq}}=\SI{4E15}{\per\square\cm}$} 
	\label{4E15_data_fit}
   \end{subfigure}
\begin{subfigure}[b]{.9\linewidth}
   \centering	\includegraphics[width=1\textwidth]{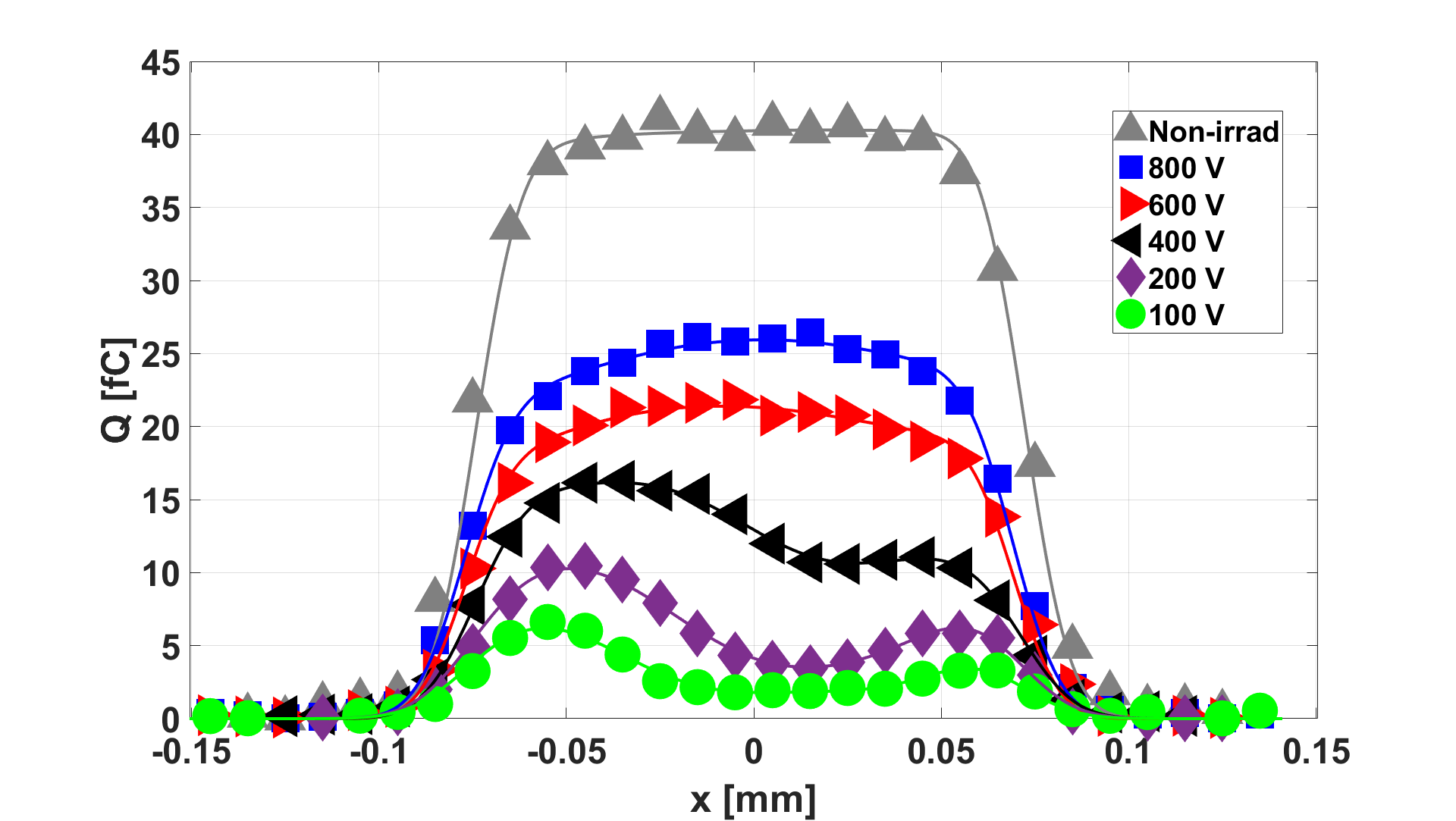} 
	\caption{$\Phi_{\text{eq}}=\SI{8E15}{\per\square\cm}$} 
	\label{8E15_data_fit}
   \end{subfigure}
      \begin{subfigure}[b]{.9\linewidth}
   \centering	\includegraphics[width=1\textwidth]{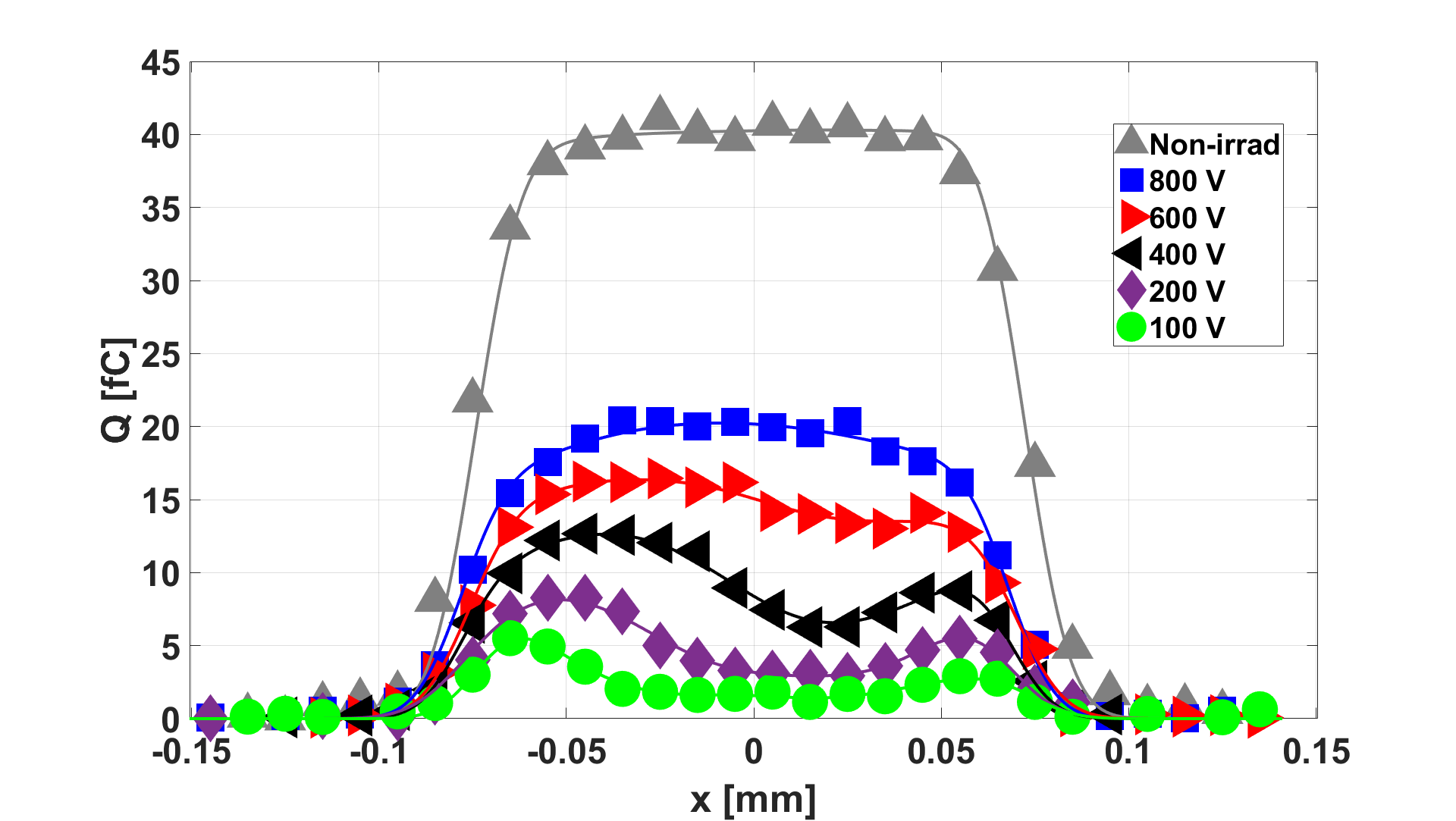} 
	\caption{$\Phi_{\text{eq}}=\SI{12E15}{\per\square\cm}$} 
	\label{12E15_data_fit}
   \end{subfigure}
   \caption{Comparison between the data and the fit results for the irradiated and non-irradiated diodes at different fluences. The fit lines are calculated charge profiles $Q_{sm}(x)$ in \cref{eq::Qunf}, after optimising the CCE spline, $\sigma$ and shift values.}
   \label{qvsx_irr_data_fit}
 \end{figure}

 \cref{qvsx_irr_data_fit} shows the comparison between data and fit results for the four irradiated and the two non-irradiated diodes. The ${CCE}_{x_i}$ profiles corrected for experimental effects are shown in \cref{ccevsx_irr}. 

\par As expected, for the non-irradiated diodes, ${CCE}_{x_i}$ profile is uniform with values around 1 and deviations not exceeding \SI{\pm 2.5}{\%} compatible with the statistical fluctuations of the data. For the irradiated diodes, the shape of the ${CCE}_{x_i}$ profiles depends on the bias voltage. At high bias voltages ($V_{\text{bias}}\leq \SI{600}{V}$), the maximum ${CCE}_{x_i}$ is close to the center of the diode, i.e.\ ($\SI{-10}{\um}<x<\SI{10}{\um}$). At low bias voltage ($V_{\text{bias}}<\SI{400}{V}$), ${CCE}_{x_i}$ is maximal at the region close to the $n^+$ implant and decreases towards the $p^+$ implant. 
\par As mentioned before, the assumed active thickness of the diode is \SI{150}{\um} for these calculations. To check this assumption, the procedure is repeated for a thickness of \SI{148}{\um} where the fit was done in the region of \SIrange{-74}{+74}{\um}. The extracted ${CCE}_{x_i}$ values for the two assumed thickness agree within \SI{1}{\%} for $x_k>-\SI{65}{\um}$. At $x_k=\SI{-65}{\um}$, the extracted $CCE$ is $\approx \SI{4}{\%}$ higher when a thickness of \SI{148}{\um} is assumed. 

\par A qualitative discussion of the $CCE$ profiles which are shown in \cref{qvsx_irr_data_fit} follows. From \cref{eq::qeqh}, it is noted that $CCE$ is a geometrical quantity. If $\lambda_e$ and $\lambda_h$ are large compared to the diode thickness, $d$, $CCE_{tot}(x) = 1$. In this case, the CCE of electrons and holes as a function of $x$ is given by: 
\begin{align*}
    CCE_e(x) &=1/2+x/d \\
    CCE_h(x) &=1/2-x/d .
\end{align*}

\par For charge-carrier absorption lengths larger than $d$, from a symmetric $CCE$ profile peaking in the centre, one can conclude that $\lambda_e \approx \lambda_h$. This appears to be the case for the irradiated diode at high voltages ($V_{\text{bias}}\leq \SI{600}{V}$). 
\par A small asymmetry, as observed at somewhat lower voltages, indicates a difference between $\lambda_e$ and $\lambda_h$: if $CCE$ is smaller at negative $x$, $\lambda_h < \lambda_e$, and $\lambda_h > \lambda_e$ for the opposite case. The second case is observed for intermediate voltages.
 
\par A flat $CCE$ minimum in the centre means that both $\lambda_e$ and $\lambda_h$ are small there, that the charges are trapped and charges entering or generated in this region will not leave it. In this case, the $CCE$ is given by the distance between the average position at which the holes are trapped and the position at which electrons are trapped divided by $d$. Thus, for the $CCE$ in the flat regions which are observed at low voltages, $CCE \approx (\lambda_e + \lambda_h)/d$.

In spite of these constraints, the determination of the position dependencies of $\lambda_e(x)$ and $\lambda_h(x)$ has not been achieved. The reasons are that \cref{eq::qeqh}, which relates $CCE$ to $\lambda_e(x)$ and $\lambda_h(x)$ is an integral equation and that from a single measured function, it appears impossible to determine two functions, $\lambda_e(x)$ and $\lambda_h(x)$.

\section{Summary and Discussion}
\par In this work, charge profiles of irradiated and non-irradiated $n^+pp^+$ diodes were measured using a \SI{5.2}{GeV} electron beam. The data have been corrected for experimental effects like finite beam resolution and transverse energy leakage. They provide a precise determination of the position dependence of the charge collection in radiation-damaged planar diodes.

\par The results of the measurements with non-irradiated diodes show that the charge collection profiles are uniform as a function of depth before irradiation. The results of the measurements with irradiated diodes at high bias voltages (${V}_{\text{bias}}\leq \SI{600}{V}$) show symmetric profiles with a peak in the centre. These results indicate a similar reduction of the charge collection lengths for electrons and holes.  

\par At intermediate bias voltages (\SIrange[]{300}{500}{\V}), the profiles are non-uniform and the $CCE$ is higher at the region close to the $n^+$ implant than the $p^+$ implant. At low bias voltages (\SIrange[]{100}{300}{V}), a uniform low-field region with low CCE is observed in the centre, and high field regions are observed around the $n^+p$ and $pp^+$ contacts. Again, the $CCE$ is higher in the $n^+p$ than in the $pp^+$ region.
\par The results of the irradiated diodes are compared with simulations using the HPTM and two Perugia models. The comparison of HPTM with the data reveals that at high bias voltages, the simulated charge collection profiles are lower than the data. This observation was found to be in agreement with previously published results when comparing the HPTM results and experimental data from pad diodes. At low bias voltages, the shape of the measured profiles is similar to the simulation. It has to be noted that these models were tuned to data with irradiation types which differ from the \SI{23}{MeV} protons presented in this paper. The precise CCE data can be used to tune and test radiation damage models, and the procedure has been demonstrated for 3 models. 
\bibliography{Ref}
\section{Acknowledgement}
The authors acknowledge support from the BMBF, the German Federal Ministry of Education and Research. This work was funded by the Deutsche Forschungsgemeinschaft (DFG, German Research Foundation) under Germany’s Excellence Strategy – EXC 2121 ‘‘Quantum Universe’’– 390833306.

The measurements leading to these results have been performed at the Test Beam Facility at DESY Hamburg (Germany), a member of the Helmholtz Association (HGF). We also thank Alexander Dierlamm and his team at the irradiation facility (KAZ) for their support.
\FloatBarrier
 
 \begin{figure}[htb]
 \begin{subfigure}[b]{.9\linewidth}
    \centering	\includegraphics[width=1\textwidth]{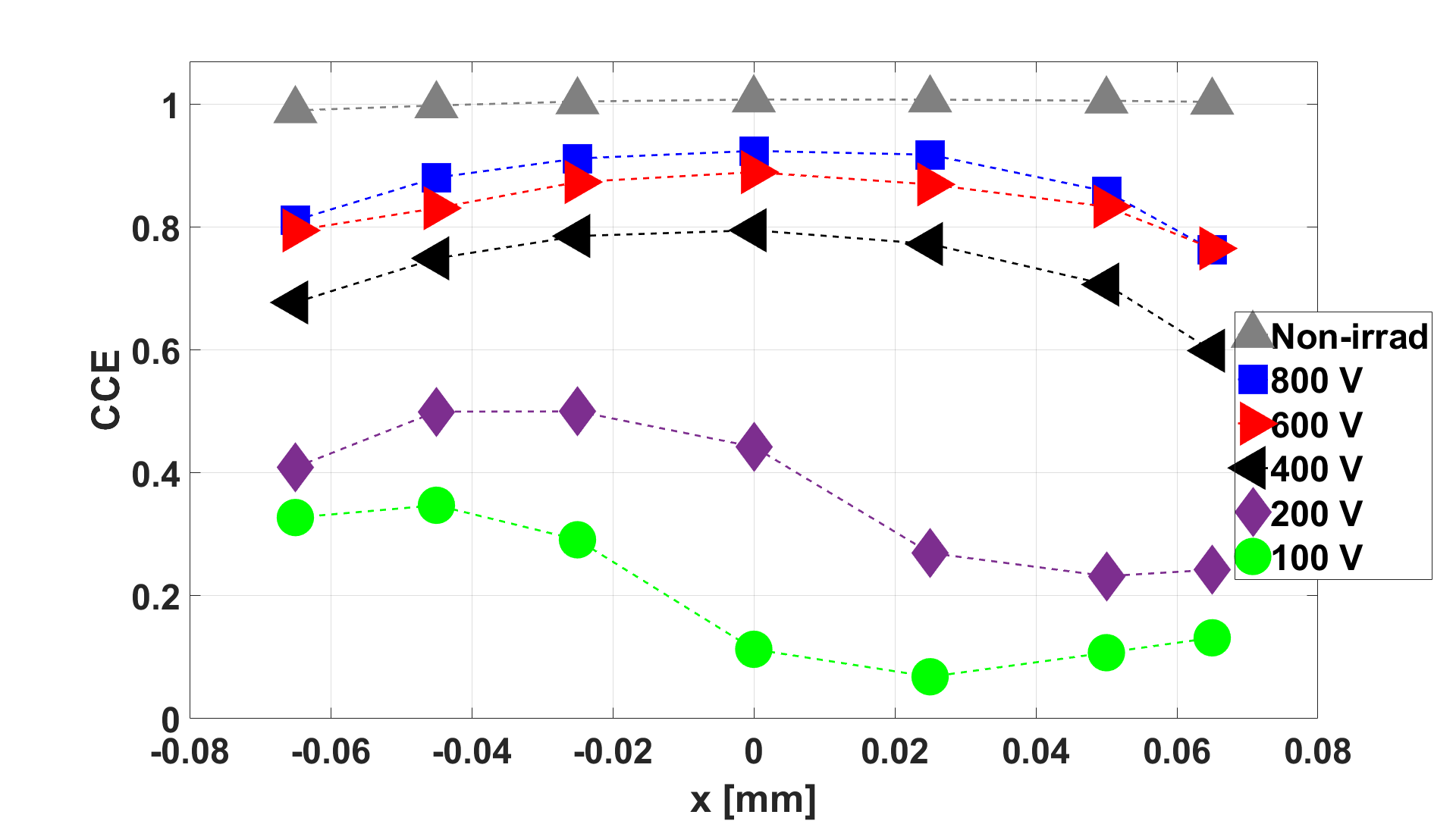} 
	\caption{$\Phi_{\text{eq}}=\SI{2E15}{\per\square\cm}$} 
     \label{ccevsx_2E15}
     \end{subfigure}
   \begin{subfigure}[b]{.9\linewidth}
   \centering	\includegraphics[width=1\textwidth]{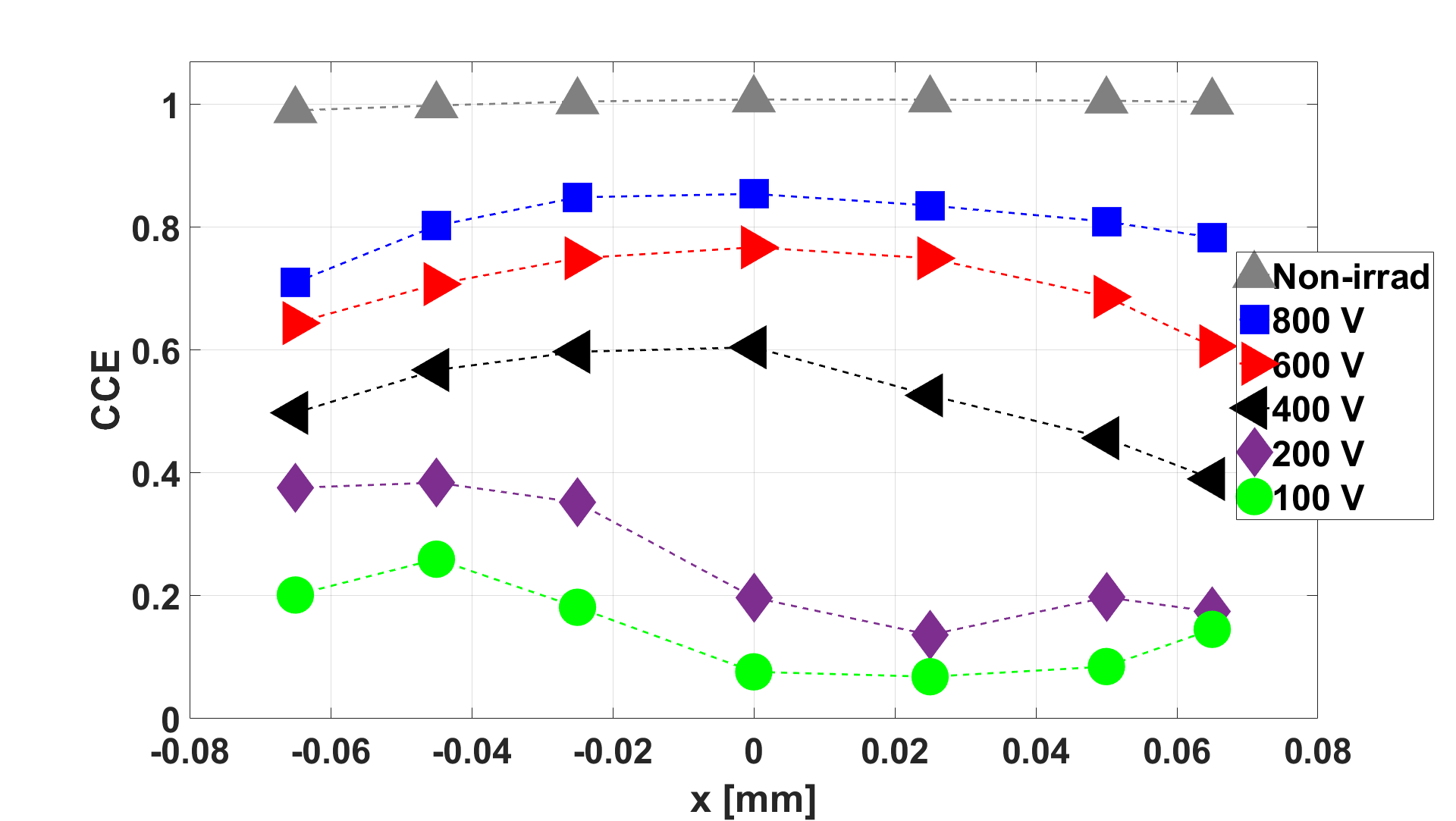} 
	\caption{$\Phi_{\text{eq}}=\SI{4E15}{\per\square\cm}$} 
	\label{ccevsx_4E15}
   \end{subfigure}
\begin{subfigure}[b]{.9\linewidth}
   \centering	\includegraphics[width=1\textwidth]{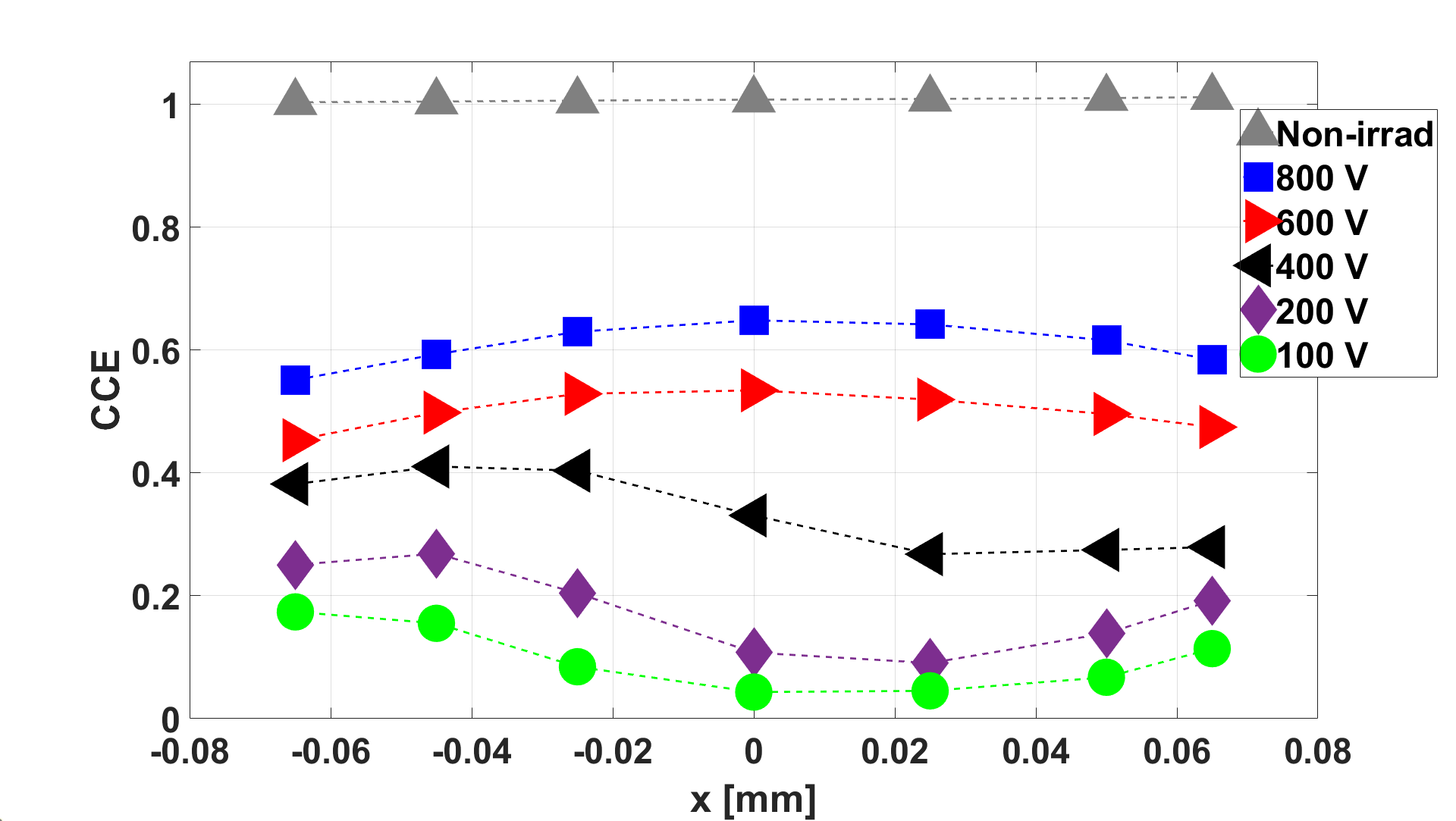} 
	\caption{$\Phi_{\text{eq}}=\SI{8E15}{\per\square\cm}$} 
	\label{ccevsx_8E15}
   \end{subfigure}
      \begin{subfigure}[b]{.9\linewidth}
   \centering	\includegraphics[width=1\textwidth]{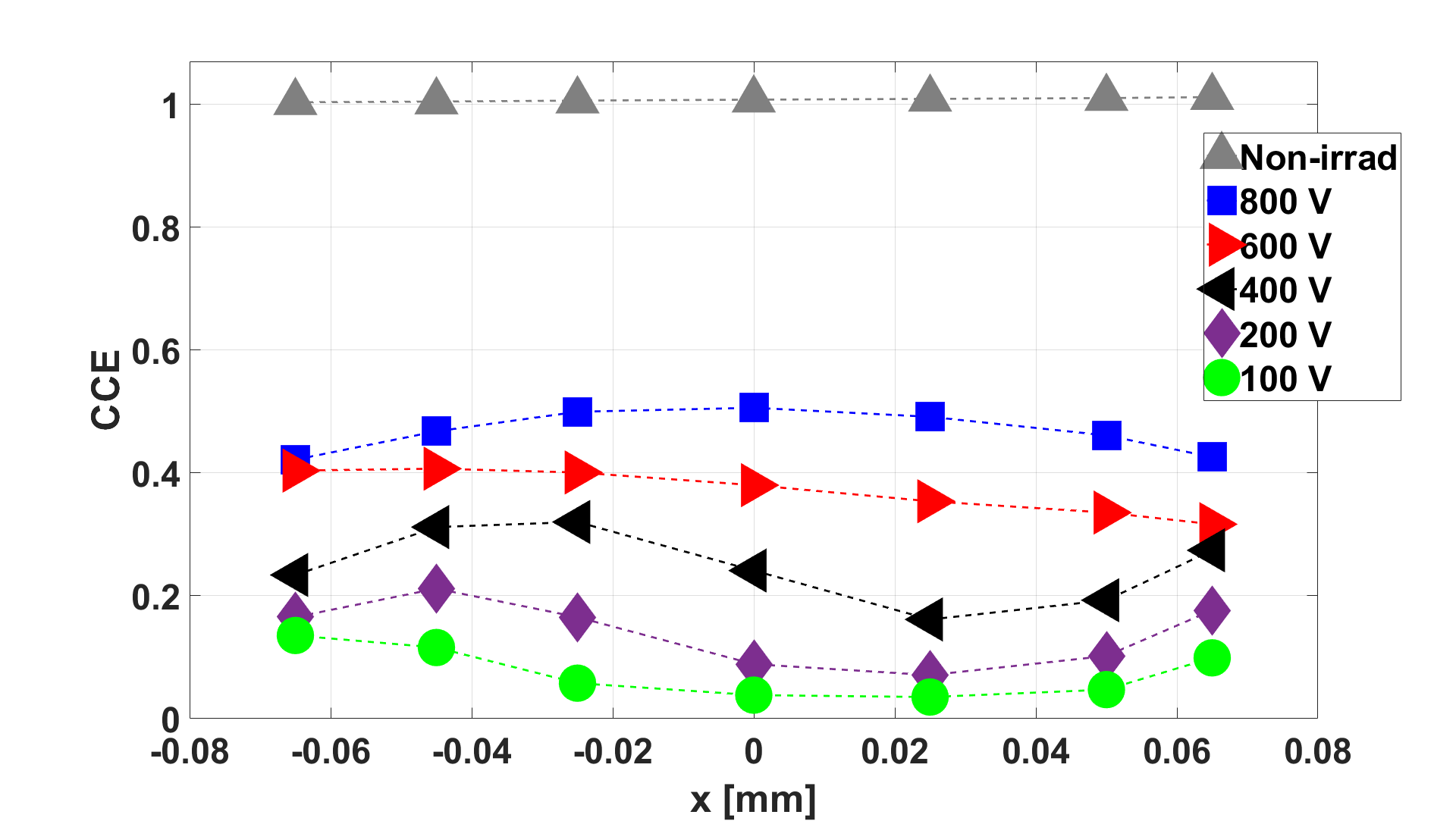} 
	\caption{$\Phi_{\text{eq}}=\SI{12E15}{\per\square\cm}$} 
	\label{ccevsx_12E15}
   \end{subfigure}
   \caption{$CCE$ profiles as a function of $x$ for the irradiated and the non-irradiated diodes. The profiles are extracted from the fits as explained in the text. }
   \label{ccevsx_irr}
 \end{figure}

\newpage
\FloatBarrier
\section{Additional Materials}
In \cref{sec::ccevsx}, ${CCE}_{x_i}$ profiles of irradiated and non-irradiated diodes were extracted and shown in \cref{ccevsx_irr}. These values are printed in \cref{table::ccevsx}.

\begin{table}[htb]
\centering
\begin{tabular}{||c c c c c c c c||} 
\hline
${CCE}_{x_i}$ & \SI{-65}{\um} & \SI{-45}{\um} & \SI{-25}{\um} & \SI{0}{\um} & \SI{25}{\um} & \SI{45}{\um} & \SI{65}{\um}\\
\hline
Non-Irrad (\SI{5.0 x 5.0}{\square \mm}) & 0.9902 &   0.9983 &   1.0047 &   1.0079 &   1.0078   & 1.0060 &   1.0041  \\
\hline
Non-Irrad (\SI{2.5 x 2.5}{\square \mm}) & 1.0035 &    1.0046    & 1.0062    & 1.0077    & 1.0090    & 1.0103    & 1.0119	\\
  \hline
Irrad ($\Phi_{\text{eq}}=\SI{2E15}{\per\square\cm}$, $V_\text{bias} = \SI{800}{V}$) & 0.8115    & 0.8808  &  0.9120 &   0.9243   & 0.9181  &  0.8583    & 0.7629  \\
  \hline  
Irrad ($\Phi_{\text{eq}}=\SI{2E15}{\per\square\cm}$, $V_\text{bias} = \SI{600}{V}$)  & 0.7948  &  0.8311    & 0.8736    & 0.8897    & 0.8698 &   0.8336 &   0.7656  \\
  \hline 
Irrad ($\Phi_{\text{eq}}=\SI{2E15}{\per\square\cm}$, $V_\text{bias} = \SI{400}{V}$) &  0.6776  &  0.7494    & 0.7857    & 0.7951    & 0.7727 &   0.7067 &   0.5990  \\
  \hline   
Irrad ($\Phi_{\text{eq}}=\SI{2E15}{\per\square\cm}$, $V_\text{bias} = \SI{200}{V}$) &  0.4091 &   0.4998    & 0.5002    & 0.4425    & 0.2697  &  0.2320 &   0.2424  \\
  \hline   

Irrad ($\Phi_{\text{eq}}=\SI{2E15}{\per\square\cm}$, $V_\text{bias} = \SI{100}{V}$) &  0.3274  &  0.3474    & 0.2912    & 0.1129    & 0.0682    & 0.1074    & 0.1316 \\
  \hline   

Irrad ($\Phi_{\text{eq}}=\SI{4E15}{\per\square\cm}$, $V_\text{bias} = \SI{800}{V}$) &  0.7104   & 0.8025    & 0.8488    & 0.8543    & 0.8354    & 0.8092    & 0.7833
 \\
  \hline   
Irrad ($\Phi_{\text{eq}}=\SI{4E15}{\per\square\cm}$, $V_\text{bias} = \SI{600}{V}$) &  0.6439  &  0.7074    & 0.7495 &    0.7673   & 0.7495 &   0.6868 &   0.6063  \\
  \hline
Irrad ($\Phi_{\text{eq}}=\SI{4E15}{\per\square\cm}$, $V_\text{bias} = \SI{400}{V}$) & 0.4979    & 0.5677    & 0.5973    & 0.6046   & 0.5263 &   0.4566 &   0.3903  \\
  \hline
Irrad ($\Phi_{\text{eq}}=\SI{4E15}{\per\square\cm}$, $V_\text{bias} = \SI{200}{V}$) &  0.3758  &  0.3843    & 0.3519    & 0.1971   & 0.1364     & 0.1975    & 0.1742 \\
  \hline
Irrad ($\Phi_{\text{eq}}=\SI{4E15}{\per\square\cm}$, $V_\text{bias} = \SI{100}{V}$) &  0.2013    & 0.2596 &   0.1813    & 0.0762   & 0.0683   & 0.0848 &   0.1454 \\
  \hline
 Irrad ($\Phi_{\text{eq}}=\SI{8E15}{\per\square\cm}$, $V_\text{bias} = \SI{800}{V}$) & 0.5515    & 0.5932 &   0.6299    & 0.6485   & 0.6418  &  0.6161 &   0.5842 \\
  \hline
 Irrad ($\Phi_{\text{eq}}=\SI{8E15}{\per\square\cm}$, $V_\text{bias} = \SI{600}{V}$) & 0.4533  &  0.4982    & 0.5288    & 0.5345   & 0.5195  &  0.4960 &   0.4747 \\
  \hline 
 Irrad ($\Phi_{\text{eq}}=\SI{8E15}{\per\square\cm}$, $V_\text{bias} = \SI{400}{V}$) & 0.3820 &   0.4105    & 0.4037    & 0.3308   & 0.2678 &   0.2747 &   0.2794 \\
  \hline
 Irrad ($\Phi_{\text{eq}}=\SI{8E15}{\per\square\cm}$, $V_\text{bias} = \SI{200}{V}$) & 0.2505  &  0.2686    & 0.2043    & 0.1075   & 0.0906 &   0.1391 &   0.1922\\
  \hline
 Irrad ($\Phi_{\text{eq}}=\SI{8E15}{\per\square\cm}$, $V_\text{bias} = \SI{100}{V}$) & 0.1738   & 0.1552    & 0.0845    & 0.0433   & 0.0453  &  0.0674  &  0.1140 \\
  \hline
 Irrad ($\Phi_{\text{eq}}=\SI{12E15}{\per\square\cm}$, $V_\text{bias} = \SI{800}{V}$) & 0.4216   & 0.4678 &   0.4996    & 0.5061   & 0.4914  &  0.4612 &   0.4258 \\
  \hline
 Irrad ($\Phi_{\text{eq}}=\SI{12E15}{\per\square\cm}$, $V_\text{bias} = \SI{600}{V}$) &  0.4038  &  0.4070 &   0.4008 &   0.3801   & 0.3539  &  0.3355  &  0.3166\\
  \hline
 Irrad ($\Phi_{\text{eq}}=\SI{12E15}{\per\square\cm}$, $V_\text{bias} = \SI{400}{V}$) & 0.2339  &  0.3119 &   0.3202    & 0.2411   & 0.1615   & 0.1928   & 0.2741  \\
  \hline
 Irrad ($\Phi_{\text{eq}}=\SI{12E15}{\per\square\cm}$, $V_\text{bias} = \SI{200}{V}$) & 0.1654  &  0.2117 &   0.1653    & 0.0887   & 0.0709  &  0.1021 &   0.1764\\
  \hline
Irrad ($\Phi_{\text{eq}}=\SI{12E15}{\per\square\cm}$, $V_\text{bias} = \SI{100}{V}$) & 0.1354  &  0.1158    & 0.0578    & 0.0383   & 0.0352  &  0.0473    & 0.0990\\
  \hline
\end{tabular}
	\caption{${CCE}_{x_i}$ profiles of the irradiated and non-irradiated diodes.} 
 \label{table::ccevsx}
\end{table}

\end{document}